

\input harvmac.tex

\Title{CLNS 92/1147, SPhT-92-054}
{\vbox{\centerline{The Quantum Double in Integrable  }
\centerline{ Quantum Field Theory} }}

\bigskip
\bigskip
\centerline{DENIS BERNARD}
\centerline{Service de Physique Th\' eorique \foot{
 Laboratoire de la Direction des Sciences de la Mati\`ere du
Commisariat \`a l'Energie Atomique.}}
\centerline{CEN-Saclay}
\centerline{91191 Gif sur Yvette, France.}
\bigskip

\centerline{ANDR\' E LECLAIR}
\medskip\centerline{Newman Laboratory}
\centerline{Cornell University}
\centerline{Ithaca, NY  14853}

\vskip .3in

Various aspects of recent works on affine quantum group symmetry
of integrable 2d quantum field theory are reviewed and further
clarified.  A geometrical meaning is given to the quantum
double, and other properties of quantum groups.  Multiplicative
presentations of the Yangian double are analyzed.

%

%
%
%
%
%
%
%
%

%
%
%
%


\def\Dep{\Delta^\prime}
\def\res{\mathop{\rm res}}
\def\u#1#2#3{{ \mu^{#1#2}_{#3} }}
\def\m#1#2#3{{ m_{#1#2}^{#3} }}
\def\vep{{\epsilon}}
\def\da{{\CD (\CA )}}
\def\dy{{\CD (\CY )}}
\def\adu{{ \CA^* }}
\def\yd{{ \CY^* }}
\def\Sm{F. A. Smirnov, `Dynamical Symmetries of Massive Integrable
Models', RIMS preprint 772, 838 (1991).}
\def\LS{A. LeClair and F. A. Smirnov,
 Int. J. Mod. Phys. A7 (1992) 2997.}
\def\FR{I. B. Frenkel and N. Yu. Reshetikhin,
proceedings of XX-th International Conference
on Differential Geometric Methods in Theoretical Physics, eds. S. Catto and
A. Rocha, World Scientific (1992).}
\def\BLnlc{D. Bernard and A. LeClair, Commun. Math. Phys. 142 (1991) 99.}
\def\FW{G. Felder and C. Wieczerkowski,
Commun. Math. Phys. 131 (1990) 125.}
\def\form{F. A. Smirnov, {\it Form Factors in Completely Integrable
Models of Quantum Field Theory}, {\it Advanced Series in Mathematical
Physics} 14, World Scientific, 1992\semi
A. N. Kirillov and F. A. Smirnov, Phys. Lett. 198B (1987) 506;
Zap. Nauch. Semin. LOMI 164 (1987) 80.}
\def\rims{B. Davies, O. Foda, M. Jimbo, T. Miwa, and A. Nakayashiki,
`Diagonalization of the XXZ Hamiltonian by Vertex Operators', RIMS
preprint (1992).}

%
\def\Luschi{M. L\"uscher, Nucl. Phys. B135 (1978) 1.}

\def\Yang{D. Bernard,
Commun. Math. Phys. 137 (1991) 191. }
\def\fssg{D. Bernard and A. LeClair,
 Phys. Lett. B247 (1990) 309.}

\def\KZ{V.G. Knizhnik and A.B. Zamolodchikov, Nucl. Phys. B297
(1984) 83.}

\def\FRT{L. D. Faddeev, N. Yu. Reshetikhin, and L. A. Takhtajan,
Algebra and Analysis 1 (1989) 178.}
\def\BPZ{A. A. Belavin, A. M. Polyakov, and A. B. Zamolodchikov,
Nucl. Phys. B241 (1984) 333.}
\def\Drinfeld{V. G. Drinfel'd, Sov. Math. Dokl. 32 (1985) 254;
Sov. Math. Dokl. 36 (1988) 212; `Quantum Groups', {\it Proceedings
of the International Congress of Mathematicians}, Berkeley, CA, 1986.}
\def\Jimbo{M. Jimbo, Lett. Math. Phys. 10 (1985) 63; Lett. Math. Phys. 11
(1986) 247; Commun. Math. Phys. 102 (1986) 537; Int. J. Mod. Phys. A4 (1989)
3759.}
\def\STF{E. K. Sklyanin, L. A. Takhtadzhyan, and L. D. Faddeev, Theor.
Math. 40 (1980) 688.}
\def\Faddeev{L. Faddeev, Les Houches Lectures 1982, Elsevier Science
Publishers (1984).}
\def\ZZ{A. B. Zamolodchikov and Al. B. Zamolodchikov, Annals
Phys. 120 (1979) 253.}
%
%
%

%
%

%
%
%
%
%

\def\ot{{\otimes}}

\def\tilde{\widetilde}
\def\bar{\overline}
\def\hat{\widehat}
\def\*{\star}
\def\[{\left[}
\def\]{\right]}
\def\({\left(}		
\def\){\right)}

%
%

\def\frac#1#2{{#1 \over #2}}
\def\inv#1{{1 \over #1}}

\def\d{\partial}
\def\der#1{{\partial \over \partial #1}}

\def\ket#1{ | #1 \rangle}

\def\rvac{\hbox{$\vert 0\rangle$}}
\def\vac{{\rvac}}
\def\lvac{\hbox{$\langle 0 \vert $}}
\def\2pi{\hbox{$2\pi i$}}

\def\dsl{\raise.15ex\hbox{/}\kern-.57em\partial}
\def\Dsl{\,\raise.15ex\hbox{/}\mkern-.13.5mu D}
%
%
\def\th{\theta}		\def\Th{\Theta}
\def\ga{\gamma}		

\def\al{\alpha}
\def\ep{\epsilon}
\def\la{\lambda}	\def\La{\Lambda}
\def\de{\delta}		\def\De{\Delta}

%
%
\def\CA{{\cal A}}	\def\CB{{\cal B}}	\def\CC{{\cal C}}
\def\CD{{\cal D}}		
		
\def\CJ{{\cal J}}		\def\CL{{\cal L}}
		
		\def\CR{{\cal R}}
\def\CS{{\cal S}}		
		
\def\CY{{\cal Y}}

\def\rvac{\hbox{$\vert 0\rangle$}}
\def\lvac{\hbox{$\langle 0 \vert $}}

\def\2pi{\hbox{$2\pi i$}}

\def\dsl{\raise.15ex\hbox{/}\kern-.57em\partial}
\def\Dsl{\,\raise.15ex\hbox{/}\mkern-.13.5mu D}
%
%
%
\font\numbers=cmss12
\font\upright=cmu10 scaled\magstep1
\def\stroke{\vrule height8pt width0.4pt depth-0.1pt}
\def\topfleck{\vrule height8pt width0.5pt depth-5.9pt}
\def\botfleck{\vrule height2pt width0.5pt depth0.1pt}
\def\Zmath{\vcenter{\hbox{\numbers\rlap{\rlap{Z}\kern 0.8pt\topfleck}\kern
2.2pt
                   \rlap Z\kern 6pt\botfleck\kern 1pt}}}
\def\Qmath{\vcenter{\hbox{\upright\rlap{\rlap{Q}\kern
                   3.8pt\stroke}\phantom{Q}}}}
\def\Nmath{\vcenter{\hbox{\upright\rlap{I}\kern 1.7pt N}}}
\def\Cmath{\vcenter{\hbox{\upright\rlap{\rlap{C}\kern
                   3.8pt\stroke}\phantom{C}}}}
\def\Rmath{\vcenter{\hbox{\upright\rlap{I}\kern 1.7pt R}}}
\def\Z{\ifmmode\Zmath\else$\Zmath$\fi}
\def\Q{\ifmmode\Qmath\else$\Qmath$\fi}
\def\N{\ifmmode\Nmath\else$\Nmath$\fi}
\def\C{\ifmmode\Cmath\else$\Cmath$\fi}
\def\R{\ifmmode\Rmath\else$\Rmath$\fi}

\Date{04/92}

\def\ot{\x}

\newsec{Introduction.}

Over the past few years it was realized that integrable quantum
field theories in 2 spacetime dimensions are characterized by
infinite dimensional Hopf algebra symmetries, i.e. affine
quantum groups.  These symmetries are generated by non-local
quantum conserved currents which may be constructed explicitly.
Previous investigations of integrable quantum field theory were
based largely on the quantum inverse scattering method
\ref\qism{\STF}\ref\fad{\Faddeev}, which deals with local integrals
of motion in involution.  The latter are useful for
diagonalizing the Hamiltonian in an algebraic Bethe ansatz scheme.
On the other hand, the non-local conserved charges generate non-abelian
symmetries, and their power resides in the rich algebraic structures
involved.  Being genuine symmetries of the field theory, they
constrain the S-matrices, form factors, and perhaps even the correlation
functions.  There is growing evidence that the affine quantum group symmetries
completely characterize the models, though much remains to be understood
towards the realization of this idea.

Let us review some previous results and the field theories we are
implicitly dealing with.   Conserved currents which generate the
quantum affine algebra $\hat{SL}_q$ in the sine-Gordon theory were
constructed in \ref\rfssg{\fssg}\ref\nlc{\BLnlc}\ref\FL{G. Felder and
A. LeClair, `Restricted Quantum Affine Symmetry of Perturbed Minimal
Conformal Models', RIMS preprint 799 and Cornell preprint CLNS 91/1099,
to appear in Int. J. Mod.
Phys.}.
This construction generalizes to the $\hat{G}$-affine Toda theories with
$\hat{G}_q$ symmetry\nlc.  At special values of the coupling where these
quantum field theories have ordinary Lie group $G$ invariance, the
quantum affine symmetry becomes the G-Yangian symmetry.  The Yangian
conserved currents were constructed in \ref\lus{\Luschi}\ref\yang{\Yang}.
Generic perturbations of rational conformal field theories are also
characterized by non-local symmetries which can be obtained as
restrictions of the above ones (see \ref\rznon{A. Zamolodchikov,
Landau Institute preprint, Sept. 1989, unpublished.}\rfssg\FL ).
The affine quantum
group invariance fixes the S-matrices up to overall scalar factors, which
in turn can be fixed from crossing symmetry, unitarity, and the
minimality assumption.  These quantum group invariant S-matrices
automatically satisfy the Yang-Baxter equation.  Thus one sees that
bootstrap principles (see e.g. \ref\rzz{\ZZ} ) are replaced by,
and actually a consequence of, the quantum affine symmetries.  The
above properties of the S-matrix are the simplest example of this idea.

In the paper \ref\rls{\LS}\ the quantum double of the symmetry was
introduced in connection with the construction of covariant field
multiplets and their form factors.  A lattice construction of
field multiplets was given in \ref\rbf{D. Bernard and G. Felder,
Nucl. Phys. B365 (1991) 98.}.
Smirnov showed that the basic axioms of the form factor bootstrap program
 may be viewed as a consequence of the affine quantum
group symmetries\ref\rsm{\Sm}.  In particular, using ideas introduced
by Frenkel and Reshetikhin\ref\rfr{\FR},  he showed how the periodicity
properties of form factors correspond to the deformed Knizhnik-Zamolodchikov
equations introduced in \rfr.
Interesting new results in this direction were obtained
in \ref\rrims{\rims}.
The form factor axioms were originally derived from bootstrap ideas\ref\rform{
\form}, so again we see that the bootstrap can be replaced with constructions
based on symmetries.

The purpose of the present work is to clarify some of the above results,
and to modestly extend them.  By introducing the general geometrical
features of non-local conserved currents, we provide a field theoretic
meaning to all of the algebraic properties of quantum groups.  Some aspects
of these
results  previously appeared in \yang\nlc\rls\rbf\ and in a
conformal field theory context in
\ref\rgs{C. G\'omez and G. Sierra, Nucl. Phys. B 352 (1991) 791.}
\ref\rfw{\FW}.
In particular we give a geometrical meaning to the quantum double.
As will be explained, the dual charges of the double are on a completely
different footing from the original non-local charges, and actually do
not correspond to conserved currents in the usual sense.  Nevertheless they
are essential for characterizing completely the S-matrix, including
scalar factors, and the braiding of non-local field multiplets.

The primary example developed in detail in this paper is
the $SL(2)$ Yangian, though the results generalize to
Yangians for other groups and to the $\hat{G}_q$ case as well.
The field theory we will be implicitly dealing
with may be formulated in a variety of ways.  As mentioned above, it
is equivalent to the sine-Gordon model at the $SL(2)$ invariant point,
i.e.  at the value of the coupling $\beta$ where the
$\cos (\beta \Phi )$ potential is marginal (conformal dimension (1,1)).
It may also be formulated as an $SL(2)$ non-abelian Thirring model,
 or chiral Gross-Neveu model.  For purposes
of our discussion, we choose to formulate it as a current-current
perturbation of the level-1 Wess-Zumino-Witten model (WZW)\ref\rkz{\KZ}:
\eqn\i{
S= S_{WZW} + \lambda \int \> d^2 x
\sum_a J^a_\mu J^a_\mu , }
where $J^a_\mu (x)$ are the local $SL(2)$ currents.
The equivalence with the previously mentioned models can be
established using bosonization. (See e.g.
\ref\rhalp{M. B. Halpern, Phys. Rev. D12 (1976) 1684.}\nlc.)
The beta function for $\lambda$ is
non-zero, and for the appropriate sign of $\lambda$, this defines
a massive field theory.  The spectrum is known to consist of an
$SL(2)$ doublet of solitons.  The S-matrix was conjectured in
\ref\rzzo{A. B. and Al. B. Zamolodchikov, Phys. Lett. 72B (1978) 481.}.

The model \i\ is characterized by the Yangian double symmetry.  This algebra
is a deformation of the full $SL(2)$ affine Lie algebra\rls.
Though the WZW theory is characterized by $SL(2)$ affine Lie
algebra symmetry, we emphasize that the Yangian double symmetry is not
to be thought of as a deformation of the affine symmetry of the conformal
WZW model.  In other words, formally setting the deformation parameter to
zero in the Yangian charges does not recover the affine Lie algebra charges of
the WZW model; i.e. the deformation parameter has nothing to do with
$\lambda$.  Rather the Yangian symmetry actually exists in the conformal
WZW model itself (actually, two copies of it), and is independent of the usual
affine Lie symmetry.  Though in the context of the WZW model itself this
Yangian symmetry has not been put to much use, its role is significant in
the perturbed massive theory since it continues to be conserved under
perturbation.  Its role is thus to organize the Hilbert space of the WZW
model in such a way that this structure is preserved under perturbation.

In this paper we will also give a different presentation of the Yangian
double than the one studied in \rls.  This is the so-called multiplicative
presentation, which was defined for the Yangian in \ref\rdrin{\Drinfeld}, and
for quantum affine algebras in \ref\rRS{N. Yu. Reshetikhin and M. A.
Semenov-Tian-Shansky, Lett. Math. Phys. 19 (1990) 133.}.  This presentation
has certain advantages in that the relations are finite in this basis.

\vfill\eject

\newsec{General Features of Quantum Group Symmetry in 2d QFT.}

In this section we will provide a quantum field theoretic meaning for
all the properties that characterize a quantum group. In this context,
a quantum group refers to an arbitrary quasitriangular Hopf algebra $\CA$.
A Hopf algebra is equipped with a multiplication map $m:\CA\ot\CA\to\CA$,
comultiplication $\De:\CA\to\CA\ot\CA$,
antipode $s:\CA\to\CA$ and counit $\ep:\CA\to C$.
In addition there is the construction
of the universal $R$-matrix, by means of the quantum double construction.
All the definitions and properties are reviewed in the appendix A, following
\rdrin\ref\rjim{\Jimbo}\ref\rfrt{\FRT}.

\subsec{Non-Local Currents and the Origin of the Comultiplication.}

We suppose we are dealing with some 2d
quantum field theory (QFT) which possesses
some conserved currents $J_a^\mu(x)$ and associated charges $Q_a$:
\eqn\IAi{ \d_\mu J_a^\mu(x)\ =\ 0
\qquad;\qquad Q_a=\int^{+\infty}_{-\infty}dx\ J^t_a(x) }
The currents $J^\mu_a(x)$ are allowed to be non-local, and consequently
to have non-trivial braiding relations among themselves and with the
other fields. Non-local fields can be described geometrically by
attaching a string from $-\infty$ to the space-time location of the
fields, where the exact shape of the string is irrelevant. This string
is completely analogous to the disorder line that defines disorder fields.
For the non-local currents in the specific models we are interested in,
the string is manifest and analogous to the Mandelstam string for the
soliton fields.
When defining the non-local currents, one must introduce a point-splitting.
This is quite natural when one realizes that the non-local currents
that have been explicitly constructed are the product of a non-local
operator (disorder field) and a local one (spin field).
So, we represent the non-local currents $J^\mu_a(x)$ as in fig.1.
\vskip 3.5 cm
$$ Fig.1 $$

Let $\Phi_i(x)$ denote a set of fields, possibly non-local.
The braiding of the non-local currents $J^\mu_a(x)$ with these
fields can be described by the equation:
\eqn\IAii{ J^\mu_a(x,t)\ \Phi_i(y,t)\ =\
\Theta^{bj}_{ai}\ \Phi_j(y,t)\ J^\mu_b(x,t)\ \qquad;\ x>y}
The above equation is implicitly time-ordered to the left.
We express the braiding in \IAii~ as an operator, as follows:
\eqn\IAiii{ J^\mu_a(x,t)\ \Phi_i(y,t)\ =\
{\hat \Theta}^b_a\({ \Phi_i(y,t)}\)\ J^\mu_b(x,t)\ \qquad;\ x>y}
where ${\hat \Th}^b_a(\Phi_i(y,t))=\Th^{bj}_{ai}\Phi_j(y,t)$.
The action of the operator $\hat \Th^a_b$ is therefore represented
graphically as in fig.2.
\vskip 3.5 cm
$$Fig.2. $$

The action of a charge on a field is defined by the expression:
\eqn\IAiv{
\hat Q_a\({\Phi_i(y)}\)\
=\ \oint_{\ga(y)}\ dz^\nu\ep_{\nu\mu}\ J^\mu_a(z)\Phi_i(y)}
where the contour $\ga(y)$ starts and ends at $-\infty$ and surrounds
the point $y$. Due to the conservation law for the currents, the exact
shape of the contour $\ga(y)$ is immaterial up to topological
obstructions. The above action of a charge on a field can be expressed
as a braided commutator by decomposing $\ga(y)$ into the contours
as in fig.3.
\vskip 3.5 cm
$$Fig. 3 $$
Therefore,
\eqn\AIv{
\hat Q_a\({\Phi_i(y)}\)\
=\ Q_a\ \Phi_i(y)\ -\ \hat \Th^b_a\({\Phi_i(y)}\)\ Q_b}
In addition to defining the action of $\hat \Th^b_a$ on a field as above,
we define the operator $\Th^b_a$ itself as a string from $-\infty$
to $+\infty$. We will denote by $\CA$ the algebra generated by the $Q_a$
and $\Th^b_a$.  The braided commutator \AIv\ may equal a new operator
if one can use the operator product expansion and close the contour on
the right hand side of \IAiv .

Let the braiding of the currents with themselves be given by:
\eqn\AIvi{\eqalign{
J^\mu_a(x)\ J^\nu_b(y)\
&=\ \hat \Th^c_a\({J^\nu_b(y)}\)\ J^\mu_c(x)\cr
&=\ R^{cd}_{ab}\ \({J^\nu_d(y)}\)\ J^\mu_c(x)\quad;\ x>y \cr }}
Then the integrated version of \AIv~ for $\Phi_j(x)=J^\mu_a(x)$ is:
\eqn\IAvii{
{\hat Q}_a\({Q_b}\)\ =\ Q_a Q_b - R^{cd}_{ab}\ Q_d Q_c}

The comultiplication arises when one considers the action of the
operators $\hat Q_a$ and $\hat \Th^b_a$ on a product of several fields.
Consider for simplicity the action on two fields:
\eqn\IAviii{
\hat Q_a\({\Phi_i(y_1)\Phi_j(y_2)}\)\ =\
\oint_{\ga(y_1;y_2)}\ dz^\nu\ep_{\nu\mu}\
J^\mu_a(z)\Phi_i(y_1)\Phi_j(y_2)}
where the contour $\ga(y_1;y_2)$ surrounds the two points $y_1,\ y_2$.
Decomposing it into the sum of two contours surrounding respectively
the points $y_1$ and $y_2$ (see figure 4), one obtains:
\eqn\IAix{
\hat Q_a\({\Phi_i(y_1)\Phi_j(y_2)}\)\ =\
\hat Q_a\({\Phi_i(y_1)}\)\Phi_j(y_2) +
\hat \Th^b_a\({\Phi_i(y_1)}\)\hat Q_b\({\Phi_j(y_2)}\) }
\vskip 3.5cm
$$Fig. 4$$
Therefore,
\eqn\IAx{\De(\hat Q_a)\ =\ \hat Q_a\ot 1 + \hat \Th^b_a\ot\hat Q_b }
Similarly, one also obtains,
\eqn\IAxi{
\De(\hat \Th^b_a)\ =\ \hat \Th^c_a\ot\hat \Th_c^b }

\subsec{Euclidean Rotations and the Origin of the Antipode and Counit.}

Define $R_\pi$ $(R_{-\pi})$ to be the Euclidean rotation in the $x-t$
plane by an angle of $\pi$ in the clockwise (anticlockwise) direction.
For $a\in\CA$, we define the antipode $s$ and skew-antipode $s'$ as follows:
\eqn\IBi{ s(a)\equiv R_\pi(a) \quad;\quad s'(a)\equiv R_{-\pi}(a)
\quad; \ \forall a\in\CA}
By construction, one has $ss'=s's=1$.
The antimultiplicative property of the antipode
$s(ab)=s(b)s(a)$ is a consequence of the fact that the rotations
$R_{\pm\pi}$ involve a time reversal.

Let us now derive the consequences of these definitions for the non-local
charges. In order to see the difference between $R_\pi$ and $R_{-\pi}$
and to realize that $R^2_\pi\not= 1$, one must remember the point-splitting
that we have introduced into the graphical notation of the non-local fields.
The rotations $R_\pi$ and $R_{-\pi}$ of $J^\mu_a(x)$ have a string
originating from $+\infty$ rather than $-\infty$: for $R_\pi(J^\mu_a(x))$
the string is above the points used in the point-splitting procedure while
for $R_{-\pi}(J^\mu_a(x))$ it is under.
Therefore $R_\pi\not= R_{-\pi}$, and $R^2_\pi(J^\mu_a(x))\not=
J^\mu_a(x)$. (See figure 5.)
\vskip 3.5 cm
$$ Fig.5  $$
This will be used at the end of section 2.6 to derive a spin-statistics
relation.

For the operators $\Th^b_a$, the rotations of $R_{\pm\pi}$ simply change
the orientation of the contour; therefore,
\eqn\IBii{
\Th^c_a\ s(\Th^b_c)\ =\ s'(\Th^b_c)\ \Th^c_a\ =\ \de^b_a }
The rotations of the operators $Q_a$ are displayed in fig.6. From
them one deduces the antipodes:
\eqn\IBiii{
s(Q_a)\ =\ - s(\Th^b_a)\ Q_b \qquad;\qquad s'(Q_a)\ =\ - Q_b\ s'(\Th^b_a) }
\vfill\eject
\topinsert
\vskip 4.0 cm
\endinsert
$$ Fig.6 $$

The counit $\ep$ of the operators $Q_a$ and $\Th^b_a$ is defined to be
the one-dimensional vacuum representation of these operators in the
QFT. Let $I$ denote the unit operator in the QFT. The counit $\ep$ may
then be defined by the formula:
\eqn\IBiv{
\vep (a) = \langle 0 | a | 0\rangle ~~~~~~\forall a\in \CA , }
or equivalently
\eqn\IBivb{
\ep(a)\ =\ {\hat a}(I) ,}
where $I$ is inserted at an arbitrary point $y$.
Indeed, it is evident from (A.5i) that the counits form a
one-dimensional representation of $\CA$.
By construction,
the counit $\ep$ satisfies:
\eqn\Iaxii{
(1\ot\ep)\De(a)=(\ep\ot 1)\De(a)=a\qquad;\qquad \forall a\in\CA}
Indeed, let $\Phi(x)$ be any quantum field, then we have:
\eqn\Iaxiii{
(1\ot\ep)\De(a)\({\Phi(x)}\) = \hat a \({\Phi(x)\ I}\)
= \hat a \({\Phi(x) }\)  }
The relation \Iaxii~ is one of the defining properties
of a Hopf algebra. Let us now derive the exact values of
the counits.  In eq. \IBiv , for the
action of $\hat Q_a$, the contour surrounding $y$ may be shrunk to zero,
therefore,
\eqn\IBv{ \ep(Q_a)=0 }
On the other hand, for the action of $\hat \Th^b_a$,
 the contour in \IBivb~ can be deformed to $-\infty$
to yield :
\eqn\IBvi{ \ep(\Th^b_a)= \de^b_a }
Finally, as a consequence of eqs. \IBii~ and \IBiii~ and of the values
of the counit, one verifies that the antipode satisfies:
\eqn\Iaxi{
m(1\ot s)\De(a)=m(s\ot 1)\De(a)= \ep(a)\qquad\qquad \forall a\in\CA}
where $m$ denotes the multiplication in $\CA$. Eq. \Iaxi~ is also one of
the compatibility relations defining a Hopf algebra.

The above properties define a Hopf algebra $\CA$, generated by
the $Q_a$ and $\Th^b_a$. One sees that the specific form of the
various operations, e.g. the comultiplication, is a consequence
of the fact that $Q_a$ is the integral of a conserved current. In
the appendix the basic properties of a Hopf algebra $\CA$ were
described using a linear basis $\{e_a\}$. There, the various
operations were formulated in terms of structure constants, i.e.:
\eqn\IBix{\eqalign{
e_a\ e_b\ &=\ \ m_{ab}^c\ e_c \cr
\De(e_a)\ &=\ \mu^{bc}_a\ e_b\otimes e_c\cr
s(e_a)\ &=\ s_a^b\ e_b \cr
\ep(e_a)\ &=\ \ep_a \cr}}
As it turns out, there always exists a change of basis to what we call
the current basis, where all the field theoretic properties described
above are valid. From the general basis $\{e_a\}$ define the generators
of the current basis as follows,
\eqn\IBxo{ Q_a = e_a-\ep_a \qquad;\qquad \Th^b_a= \mu^{cb}_a\ e_c}
Using the properties of the generators $e_a$ in the appendix, one
verifies the relations \IAx~ \IAxi~ (without the hat) and
\IBii~ \IBiii~ \IBv~ \IBvi.  We will henceforth use the basis
$\{e_a\}$ and $\{Q_a,\Th^b_a\}$ interchangeably for the algebra
$\CA$.

The information about the multiplication maps for $\CA$ is contained in
the relation \IAvii. For reasons that will become apparent in the next
subsection, we make this connection by introducing an adjoint action
of the algebra $\CA$ on itself. For $g\in\CA$, express $\De(g)$ as
\eqn\IBx{\De(g)\ =\ \sum_i g_i\ot g^i\qquad;\quad g_i,g^i\in\CA}
and define the adjoint action of $g$ on $h$ as:
\eqn\IBxi{ad_g(h)\ =\ \sum_i\ g_i \> h \> s(g^i )}
This adjoint action enjoys the property, for $g_1,\ g_2\in\CA$,
\eqn\IBxii{
ad_{g_1}\({ad_{g_2}(h)}\)\ =\ ad_{g_1g_2}(h), \qquad h\in\CA}
which is a consequence of the homomorphism property of the comultiplication
and of the anti-multiplicative property of the antipode. In particular,
one finds
\eqn\IBxiii{ ad_{e_a}(e_b)\ =\ {\tilde f}^c_{ab}\ e_c }
where
\eqn\IBxiv{ {\tilde f}^c_{ab}\ =\ \mu^{ij}_a\ s_j^l\ m_{ib}^k\ m^c_{kl} }

In the current basis, one finds,
\eqn\IBxv{\eqalign{
 ad_{\Th^b_a}(Q_c)\ &=\ R^{bd}_{ac}\ Q_d\cr
ad_{Q_a}(Q_b)\ &=\ Q_aQ_b-R^{cd}_{ab}Q_dQ_c\ =\ f^c_{ab}\ Q_c\cr}}
where
\eqn\IBxvii{\eqalign{
R^{bd}_{ac}\ &=\ \mu^{ib}_a\ {\tilde f}^d_{ic}\
=\ \mu_a^{ib}\mu^{jk}_i\ s_k^l\ m_{jb}^n\ m^d_{kn}\cr
f^c_{ab}\ &=\ {\tilde f}^c_{ab}-\ep_a\de^c_b\cr}}
One may clearly formulate all the general Hopf algebra properties
in the current basis.
It is clear that \IBxv~ contains the information of the multiplication
maps. We have already explained how the l.h.s. of \IBxv~ arises as a braided
commutator. As explained above, the r.h.s. of \IBxv~ arises when one
can use the operator product expansion to compute the action of
the charges on the currents.

\subsec{Adjoint Action and Field Representations.}

In the situation where one is dealing with ordinary Lie algebra
symmetry in QFT generated by local currents (which just corresponds
to $\Th^b_a=\de^b_a$ above) one finds that fields can be organized into
irreducible representations of the Lie algebra. The action of a
conserved charge an a field is simply the commutator of the charge with
the field. Starting from a highest weight field, one can complete the
representation by acting with appropriate raising or lowering
operators. That the fields related via this action necessarily form
a representation of the Lie algebra is ensured by the Jacobi identity.

The above notions may be generalized to the situation of an arbitrary
Hopf algebra symmetry. Several different definitions of the adjoint action
of a charge on a field are possible; for the case of Lie algebra symmetry
they are equivalent. Let $g\in\CA$, with $\De(g)=\sum g_i\otimes g^i$.
We define two different adjoint actions of $g$ on a field $\Phi(x)$
as follows:
\eqn\ICi{\eqalign{
ad_g\({\Phi(x)}\)\ &=\ \sum_i\ g_i\Phi(x) s(g^i) \cr
ad'_g\({\Phi(x)}\)\ &=\ \sum_i\ g^i\Phi(x) s'(g_i) \cr}}
Using the Hopf algebra properties outlined in the appendix, one
shows that the adjoint actions $ad$ and $ad'$ enjoy the following
properties:
\eqn\ICii{\eqalign{
ad_{g_1}\({ad_{g_2}(\Phi(x))}\)\ &=\ ad_{g_1g_2}\({\Phi(x)}\)\cr
ad'_{g_1}\({ad'_{g_2}(\Phi(x))}\)\ &=\ ad'_{g_1g_2}\({\Phi(x)}\)\cr}}
Furthermore,
\eqn\ICiii{\eqalign{
ad_g(\Phi_1(x_1)\Phi_2(x_2))\ &=\
\sum_i\ ad_{g_i}(\Phi_1(x_1)\ ad_{g^i}(\Phi_2(x_2))\cr
ad'_g(\Phi_1(x_1)\Phi_2(x_2))\
&=\ \sum_i\ ad'_{g^i}(\Phi_1(x_1)\ ad'_{g_i}(\Phi_2(x_2))\cr}}
The adjoint actions \ICii~ are ``left" actions and one also has the
possibility of defining two ``right" actions:
\eqn\ICiv{\eqalign{
ad^R_g\({\Phi(x)}\)\ &=\ \sum_i\ s(g_i)\Phi(x) g^i \cr
ad'^{~R}_g\({\Phi(x)}\)\ &=\ \sum_i\ s'(g^i)\Phi(x) g_i \cr}}
These actions possess the analogous properties:
\eqn\ICv{\eqalign{
ad^R_{g_1}\({ad^R_{g_2}(\Phi(x))}\)\ &=\ ad^R_{g_2g_1}\({\Phi(x)}\)\cr
ad'^{~R}_{g_1}\({ad'^{~R}_{g_2}(\Phi(x))}\)\
&=\ ad'^{~R}_{g_2g_1}\({\Phi(x)}\)\cr}}
Henceforth, we will only be using the left actions.

For the generators $Q_a$ and $\Th^b_a$, the adjoint actions \ICi~ read:
\eqn\ICvi{\eqalign{
ad_{\Th^b_a}(\Phi(x))\ &=\ \Th^c_a\ \Phi(x)\ s(\Th^b_c)\cr
ad_{Q_a}(\Phi(x))\ &=\ Q_a\Phi(x)-ad_{\Th^b_a}(\Phi(x))Q_b\cr}}
and
\eqn\ICvii{\eqalign{
ad'_{\Th^b_a}(\Phi(x))\ &=\ \Th^b_c\Phi(x) s'(\Th^c_a)\cr
ad'_{Q_a}(\Phi(x))\ &=\ \Phi(x)s'(Q_a)+Q_b\Phi(x)s'(\Th^b_a)\cr}}
Using the graphical representations for $Q_a$ and $\Th^b_a$ and their
antipodes, one can formulate the equations \ICvi~ and \ICvii. In this way it
is clear that the graphical formulation of the r.h.s. of \ICvi~ is the
same as displayed for the operators $Q_a$ and $\Th^b_a$
in fig.2 and fig.3 respectively. Therefore, we have an equivalence
between the formula \ICvi~ and the actions on fields ${\hat Q}_a$
and ${\hat \Th}^b_a$ defined in section 2.1.:
\eqn\ICix{\eqalign{
ad_{Q_a}(\Phi(x))\ &=\ {\hat Q}_a(\Phi(x)) \cr
ad_{\Th^b_a}(\Phi(x))\ &=\ {\hat \Th}^b_a(\Phi(x)) \cr}}
The graphical formulation of eq. \ICvii~ is formulated similarly,
and one finds that the contours originate at $+\infty$ rather that $-\infty$.
(See figure 7.)
\vskip 4.0cm
$$Fig. 7$$

\def\too{{ \mathop{\longrightarrow} }}
\def\c{\cdot}
\def\PL{\Phi_\La(x)}
\def\PLp{\Phi_\La'(x)}
A  definition of a highest weight field $\Phi_\la(x)$ will
be given in the next section. Starting from such a highest weight
field, repeated $ad$ or $ad'$ action yields a tower of its
descendent fields $\Phi_\La$ and $\Phi_\La'$:
\eqn\ICx{
\Phi_\la(x)\  \too^{ad}     \ \Phi_\La(x)
\quad ;\quad
\Phi_\la(x)\  \too^{ad'} \ \Phi_\La'(x) }
In general, the fields $\Phi_\La(x)$ or $\Phi_\La'(x)$ will be
non-local, since the charges are non-local. The fields $\Phi_\La(x)$
and $\Phi_\La'(x)$ span a vector space $\La$. We denote individual
fields in this vector space as $\Phi_v(x)$ and $\Phi_v'(x)$, $v\in\La$.
Due to the properties \ICii, the fields $\PL$ and $\PLp$
necessarily form a representation of $\CA$:
\eqn\ICxi{\eqalign{
ad_g\({\Phi_v(x)}\)\ &=\ \rho_\La(g)^w_v\ \Phi_w(x)\
\equiv\ \Phi_{g\cdot v}(x)\cr
ad'_g\({\Phi'_v(x)}\)\ &=\ \rho'_\La(g)^w_v\ \Phi'_w(x)\
\equiv\ \Phi'_{g\c v}(x)\cr}}
In general, for the affine quantum groups the fields $\Phi_\Lambda$ will
form infinite dimensional Verma module representations.

Using the Hopf algebra properties one can show that for $g\in \CA$:
\eqn\ICxii{\eqalign{
g\ \Phi(x)\ &=\ ad_{g_i}\({\Phi(x)}\)\ g^i \cr
g\ \Phi(x)\ &=\ ad'_{g^i}\({\Phi(x)}\)\ g_i \cr
\Phi(x)\ s(g)\ &=\ s(g_i)\ ad_{g^i}\({\Phi(x)}\) \cr
\Phi(x)\ s'(g)\ &=\ s'(g^i)\ ad'_{g_i}\({\Phi(x)}\) \cr}}
where $g_i$ and $g^i$ are defined in eq. \IBx. By comparing the graphical
formulation for $ad_g(\Phi(x))$ and $ad'_g(\Phi(x))$ when $g=Q_a$ or
$\Th^b_a$, one sees that $\PL$ and $\PLp$ are simply related by a rotation:
\eqn\ICxiii{ \Phi'_\La(0)\ =\ R_{-\pi}\({\Phi_\La(0)}\) }
Indeed, eq. \ICxiii~ ensures that if $\PL$ is $ad$-covariant, then $\PLp$
is $ad^\prime$-covariant. The proof is straightforward:
\eqn\ICxiv{\eqalign{
R_{-\pi}\({ ad_g(\Phi_v(0))}\)\ &=\ R_{-\pi}\({g_i\Phi_v(0)s(g^i)}\)\cr
	&=\ s's(g^i)\ R_{-\pi}(\Phi_v(0))\ s'(g_i)\cr
	&=\ ad'_g\({\Phi'_v(0)}\) \cr}}
In the second step we have used the fact that $R_{-\pi}$ involves a time
reversal, and also the relation $ss'=1$.

\subsec{The Geometrical Meaning of the Quantum Double.}

As before, we let $\{e_a\}$ denote a general linear basis for $\CA$.
The charges  $e_a$ act on a field via the adjoint actions defined above:
\eqn\IDi{\eqalign{
ad_{e_a}\({\Phi_v(x)}\)\
&=\ \mu^{bc}_a\ e_b\Phi_v(x)s(e_c)\equiv\Phi_{e_a \c v}(x)\cr
ad'_{e_a}\({\Phi'_v(x)}\)\
&=\ \mu^{bc}_a\ e_c\Phi'_v(x)s'(e_b)\equiv\Phi'_{e_a \c v}(x)\cr}}

In seeking an operator formulation of the rotations $R_{\pm\pi}$, one is led
to define a different dual Hopf algebra $\CA^*$ whose properties can be
derived from its operational definition. We let $\{e^a\}$ denote a linear
basis for $\CA^*$. We will give two different, but closely related
definitions of the action of $e^a$ on fields, one being more suitable
than the other, depending on which property of $e^a$ one is interested in.
We denote the action of $e^a$ on a field as:
\eqn\IDii{ e^a\({\Phi_v(x)}\)\ \equiv\ \Phi_{e^a \c v}(x) }
and similarly for $e^a\({\Phi'_v(x)}\)$.
We emphasize that we do not assume the elements $e^a$ can be defined in
the QFT as integrals of conserved currents, the way the elements of
$\CA$ were.  Rather we provide a set of `instructions' for defining
{\it only}
the action of $e^a$ on a field, and then proceed to derive its algebraic
properties.

The first definition of $e^a(\Phi_v(x))$ is with respect to the
braiding of fields:
\eqn\IDiii{\eqalign{
\Phi_{v_1}(x_1)\ \Phi_{v_2}(x_2)\ &=\ \sum_a
\Phi_{e_a \c v_2}(x_2)\ \Phi_{e^a \c v_1}(x_1)\ \quad;\quad x_1>x_2\cr
\Phi'_{v_1}(x_1)\ \Phi'_{v_2}(x_2)\ &=\ \sum_a
\Phi'_{e^a \c v_2}(x_2)\ \Phi'_{e_a \c v_1}(x_1)\ \quad;\quad x_1>x_2\cr}}
In the above braiding relations, the fields $\Phi_{v_1}(x)$
and $\Phi_{v_2}(x)$ may be part of two different representations.
The algebraic relations among the elements of $\CA$ and $\CA^*$ arises
when one considers the $\CA$-covariance of \IDiii. Applying $ad_{e_a}$ to both
sides of eq. \IDiii~ and using eq. \ICiii~ one obtains:
\eqn\IDiv{\eqalign{ ad_{e_a}\({ {\rm L.H.S.} }\)\
&= \mu^{bc}_a\ \Phi_{e_b \c v_1}(x_1)\ \Phi_{e_c \c v_2}(x_2)\cr
&= \mu^{bc}_a\ \Phi_{e_ne_c \c v_2}(x_2)\ \Phi_{e^ne_b \c v_1}(x_1)\cr}}
whereas
\eqn\IDv{ad_{e_a}\({ {\rm R.H.S.} }\)\
= \mu^{bc}_a\ \Phi_{e_be_n \c v_2}(x_2)\ \Phi_{e_ce^n \c v_1}(x_1)}
Equating \IDiv~ and \IDv~ one obtains :
\eqn\IDvi{ \mu^{bc}_a\ m^d_{nc}\ e^n\ e_b\ =\
\mu^{bc}_a\ m^d_{bn}\ e_c\ e^n\ }
The $\CA$-covariance of the braiding relations for the $ad'$-covariant
fields is also ensured by eq.\IDvi. Eq. \IDvi~ gives the commutation
relations between the $\{e_a\}$ and the $\{e^b\}$.

The multiplication map in $\CA^*$ can be derived by requiring that the
algebra \IDiii~ is associative when one braids a product of three fields.
This requirement is equivalent to the Yang-Baxter equation (A.9), which is
satisfied as a consequence of \IDvi~ if the following additional relations
are imposed:
\eqn\IDvii{ e^a\ e^b\ =\ \mu^{ba}_c\ e^c .}
Thus, the braiding of field multiplets is given by the universal
$\CR$-matrix evaluated in the representations $\rho_{\La_1} , \rho_{\La_2}$:
\eqn\braid{
\Phi_{\La_1} (x_1) \> \Phi_{\La_2} (x_2)
=\CR_{\rho_{\La_2} , \rho_{\La_1} } ~ \Phi_{\La_2} (x_2)
\> \Phi_{\La_1} (x_1) ~~;    ~~~~x_1 > x_2 . }

The second definition of the action of $e^a$ on a field involves directly
the rotation $R_{\pm\pi}$, and is conceptually more fundamental,
\eqn\IDviii{\eqalign{
R_\pi\({\Phi'_v(0)}\)\ &=\ \Phi_v(0)\ =\ e^a\({\Phi'_v(0)}\)\ e_a\cr
R_{-\pi}\({\Phi_v(0)}\)\ &=\ \Phi'_v(0)\ =\ e^a\({\Phi_v(0)}\)\ s'(e_a)\cr}}
One may check that the formulas \IDviii~ are covariant if \IDvi~ is satisfied.
I.e. the formulas \IDviii~ are compatible with the covariance properties
\IDi~ of both field multiplets $\Phi_v$ and $\Phi'_v$. Let us
consider the first of the relations \IDviii.
To prove the covariance,  we begin with the following relations (eq. \ICxii):
\eqn\IDix{\eqalign{
e_a\ \Phi_v\ &=\ \mu^{bc}_a\ \Phi_{e_b \c v}\ e_c \cr
e_a\ \Phi'_v\ &=\ \mu^{bc}_a\ \Phi'_{e_c \c v}\ e_b . \cr}}
The latter imply
\eqn\IDx{
e_a\ \Phi_v\ =\ e_a\ \Phi'_{e^n \c v}\ e_n\ =\
\mu^{bc}_a\ m^d_{bn}\ \Phi'_{e_c e^n \c v}\ e_d}
On the other hand, one also has,
\eqn\IDxi{
\mu^{bc}_a\ \Phi_{e_b \c v}\ e_c\ =\ \mu^{bc}_a\ \Phi'_{e^ne_b \c v}\ e_ne_c\
=\
\mu^{bc}_a\ m^d_{nc}\ \Phi'_{e^n e_b \c v}\ e_d}
Equating \IDx~ and \IDxi~ one concludes that \IDviii~ is covariant if
\IDvi~ is satisfied. In the same way one verifies that the second
relation \IDviii~ is also covariant.

The definition \IDviii~ of $e^a(\Phi(x))$ allows
one to {\it derive}  the comultiplication $\De(e^a)$. This comultiplication
arises when one considers the action of $e^a$ on a product of two fields.
The definition \IDviii~ extended to such a product reads:
\eqn\IDxii{\eqalign{
\Phi_{v_1}(x_1)\ \Phi_{v_2}(x_2)\
&=\ e^a\({ R_{-\pi}\( \Phi_{v_1}(x_1) \Phi_{v_2}(x_2)\)}\)\ e_a \cr
&=\ e^a\({ \Phi'_{v_2}(x_2) \Phi'_{v_1}(x_1)}\)\ e_a \cr}}
Let $\De'(e^a)=L^a_{bc} e^c\ot e^b$. Then,
\eqn\IDxiii{\eqalign{
e^a\({ \Phi'_{v_2}(x_2) \Phi'_{v_1}(x_1)}\)\ e_a
&=\ e^a\({ \Phi'_{e^n \c v_1}(x_1) \Phi'_{e_n \c v_2}(x_2)}\)\ e_a \cr
&=\ L^a_{kl}\ \Phi'_{e^le^n \c v_1}(x_1) \Phi'_{e^ke_n \c v_2}(x_2)\ e_a \cr
&=\ L^a_{kl}\ \mu^{nl}_m\
\Phi'_{e^m \c v_1}(x_1) \Phi'_{e^ke_n \c v_2}(x_2)\ e_a \cr}}
On the other hand,
\eqn\IDxiv{\eqalign{
\Phi_{v_1}(x_1)\ \Phi_{v_2}(x_2)\
&=\ \Phi'_{e^k \c v_1}(x_1)\ e_k\ \Phi'_{e^j \c v_2}(x_2)\ e_j\cr
&=\ \mu^{sr}_k\ \Phi'_{e^k \c v_1}(x_1)\ \ \Phi'_{e_re^j \c v_2}
(x_2)\ e_se_j\cr
&=\ \mu^{sr}_k m^n_{sj}\ \Phi'_{e^k \c v_1}(x_1)\ \
\Phi'_{e_re^j \c v_2}(x_2)\ e_n\cr}}
Equating \IDxiii~ and \IDxiv~ requires
\eqn\IDxv{ L^i_{sr}\ \mu^{jr}_k\ e^se_j\ =\
\mu^{sr}_k\ m^i_{sj}\ e_re^j }
Comparing with eq. \IDvi, one concludes :
\eqn\IDxvi{ \De(e^a)\ =\ m^a_{bc}\ e^b\ot e^c .}
The same comultiplication can be derived from the second relation \IDviii.
The antipode and counit in $\CA^*$ can be fixed to be consistent with
the above multiplication and comultiplication, as in the
appendix: $\ep(e^a)=\ep^a$, $s(e^a)=(s')^a_b e^b$. One also has
$\ep_a e^a=1$. The relations \IBix~ and \IDvi~ \IDvii~ \IDxvi~ define the
double Hopf algebra $\CD(\CA)$.

We are now in position to define more precisely a highest weight
field.
Consider highest weight fields that are defined to be
annihilated by the elements of $\CA^*$.
Recall the counit was defined to be the one-dimensional vacuum
representation and we take the same definition for $\CA^*$.
The statement that $e^a$ annihilate highest weight fields amounts
to demanding that $e^a$ on $\Phi_\la(x)$ gives this one-dimensional
representation:
\eqn\IDxvii{ e^a\({\Phi_\la(x)}\)\ =\ \ep^a\ \Phi_\la(x)
\quad;\quad
e^a\({\Phi'_\la(x)}\)\ =\ \ep^a\ \Phi'_\la(x)}
One can show using the properties in appendix A that $\ep^a$ indeed defines
a one-dimensional representation of $\adu$: $\ep^a \ep^b = \mu^{ba}_c \ep^c$.
This definition is consistent with the
fact that  for highest weight fields $\Phi_\la$ must equal $\Phi'_\la$;
from \IDviii~ one has:
\eqn\IDxviii{ \Phi_\la\ =\ e^a(\Phi'_\la) e_a =
\Phi'_\la\ \ep^ae_a\ =\ \Phi'_\la}
Highest weight fields defined this way
are necessarily local with respect to each
other; from \IDiii\ one has:
\eqn\IDxix{\eqalign{
\Phi_{\la_1}(x_1)\ \Phi_{\la_2}(x_2)\
&=\ \Phi_{e_n.\la_2}(x_2)\ \Phi_{e^n.\la_1}(x_1)\ \cr
&=\ \ep^n\ \Phi_{e_n.\la_2}(x_2)\ \Phi_{\la_1}(x_1)\ \cr
&=\ \Phi_{\la_2}(x_2)\ \Phi_{\la_1}(x_1)\ \cr}}
In order to characterize all the highest weight fields, including
e.g. the soliton fields that create massive particles, one must extend
the above definition of highest weight field to include non-local ones.
More generally then, one can define highest weight fields to form a
finite dimensional representation of some abelian subalgebra of the
double.  For example, in the case of the Yangian, the soliton fields
can be taken as highest weight, transforming in the spinor representation
of the $SL(2)$ subalgebra.

Finally, it is possible to establish directly the equivalence
of the two above definitions of the action of $e^a$ on a field by
actually deriving the braiding relation \IDiii~ from \IDviii.
Since $\Phi_\La$ and $\Phi'_\La$ are related by a rotation $R_\pi$, whereas
the strings that characterize the non-locality of $\Phi_\La$ originate
at $-\infty$, those for $\Phi'_\La$ originate at $+\infty$. Therefore,
there is no obstruction to deforming the contour in the product of
$\Phi_{v_1}(x_1)\Phi'_{v_2}(x_2)$ for $x_1<x_2$; thus
\eqn\IDxx{
\Phi_{v_1}(x_1)\Phi'_{v_2}(x_2)\ =\
\Phi'_{v_2}(x_2)\Phi_{v_1}(x_1)\quad;\ x_1<x_2 }
Inserting \IDviii~ into \IDxx~ one has:
\eqn\IDxxi{
\Phi_{v_1}(x_1) \> s(e_a ) \> \Phi_{e^a \c v_2}(x_2)\ =\
s(e_a)\ \Phi_{e^a \c v_2}(x_2) \> \Phi_{v_1} (x_1 ) }
Replacing $v_2$ by $e^b \c v_2$ and using \IDvii, eq. \IDxxi~ becomes:
\eqn\IDxxii{
\mu^{ba}_c\ \Phi_{v_1}(x_1) s(e_a)\Phi_{e^c \c v_2}(x_2)\ =\
\mu^{ba}_c\ s(e_a)\ \Phi_{e^c \c v_2}(x_2)\Phi_{v_1} (x_1) }
Multiplying both sides of \IDxxii~ by $e_b$ and summing over $b$ one obtains
\IDiii .

\subsec{Adjoint Representations of the Double and the Current Multiplet.}

We began section 2.1 by postulating non-local
conserved currents $J^\mu_a(x)$ for the charges $Q_a\in\CA$. It is
natural to suppose that the complete set of these non-local currents
comprise a representation of the double algebra $\CD(\CA)$, which
we will refer to as the current representation $\rho_{{}_J}$.

For the elements of $\CA$, the current representation is defined
as arising through the adjoint action:
\eqn\IEi{ ad_{e_a}\({ Q_b}\)\ =\ {\tilde f}^c_{ab}\ Q_c
\equiv\ \rho_{{}_J}(e_a)^c_b\ Q_c.}
The local form \IEi~ can be taken as:
\eqn\IEii{ ad_{e_a}\({ J^\mu_b(x)}\)\ =\ {\tilde f}^c_{ab}\ J^\mu_c(x)}
The current representation of elements of the dual algebra $\CA^*$ on
the non-local currents may be computed directly from the definition
\IDviii. One has:
\eqn\IEiii{\eqalign{
R_{-\pi}(J^\mu_a(x))\ &=\ e^n\({J^\mu_a(x)}\)\ s'(e_n)\cr
R_\pi(J^\mu_a(x))\ &=\ s(e_n) e^n\({J^\mu_a(x)}\)\cr}}
The field $J'^\mu_a(x)$ is related to $J^\mu_a(x)$ by a rotation:
$J'^\mu_a(x)=R_{-\pi}(J^\mu_a(x))$. Thus the string for $J'^\mu_a(x)$
originates at $+\infty$ and ends at $x$. Deforming the string such that
it first goes to $-\infty$ before continuing to $x$, one finds:
\eqn\IEiv{ J'^\mu_a(x)\ =\ J^\mu_b(x)\ s'(\Th^b_a) .}
Comparing \IEiv~ with \IEiii~ and using \IBxo, one finds:
\eqn\IEv{ e^n\({ J^\mu_a(x)}\)\ =\ \mu^{nb}_a\ J^\mu_b(x)
\equiv \rho_J (e^n)_a^b \> J_b^\mu (x).}
The current representation defined in eqs. \IEii~ and
\IEv~ indeed satisfies the complete set of relations of $\CD(\CA)$.
The proof is given in appendix B.

An important aspect of the development in section 2.1 was the interpretation
of $\Th^b_a$ as a braiding operator. The fact that the currents are in
the current representation, along with the general braiding relations \IDiii,
leads directly to a verification of the above interpretation of $\Th^b_a$.
Indeed we have:
\eqn\IEvi{\eqalign{
J^\mu_a(x)\ \Phi_\La(y)\
&=\ \Phi_{e_d \cdot\La}(y)\ e^d\({J^\mu_a(x)}\)
\qquad;\ x>y\cr
&=\ \mu^{db}_a\ \Phi_{e_d \cdot\La}(y)\ J^\mu_b(x)\cr
&=\ ad_{\Th^b_a}\({\Phi_{\La}(y)}\)\ J^\mu_b(x) . \cr}}
This illustrates once more the relation between braiding and the
comultiplication.

Finally, we use the formulas \IEiii~ to derive a spin-statistics relation.
The spin is defined through a rotation of $2\pi$:
$\CS pin(J_a^\mu) \equiv R^2_\pi(J^\mu_a)$. Using eq. \IEiii~ one deduces,
\eqn\IEvii{ R^2_\pi(J^\mu_a)\ =\ ad_{s(\Th^d_a)} \( J^\mu_d \) .}
More explicitly, let us introduce
the inverse braiding matrix ${\bar R }^{cd}_{ab} = \mu^{ib}_a s^j_i
\tilde{f}^d_{jc} $ characterized by
${\bar R}^{bm}_{kc} R^{kd}_{am}=\de^d_c\de^b_a$. Then,
\eqn\IEviii{
\CS pin(J_a^\mu)\ \equiv \CS^b_a\ J^\mu_b\ =\ {\bar R}^{db}_{ad}\ J^\mu_b.}
Equivalently, the spin matrix is characterized by:
\eqn\IEix{
\CS^m_n\ R^{nb}_{am}\ =\ \de^b_a .}
Eq. \IEix~ is represented in figure 8.
\vskip 3.0cm
$$Fig. 8$$
For currents with definite Lorentz spin, $\CS^b_a$ is diagonal:
\eqn\espin{
R_{\pi}^2 (J_a (x) ) = e^{2\pi i s_a} \> J_a (x), }
where $s_a$ is the Lorentz spin of the field $J_a (x)$ ${\rm mod}~ 1$.

\subsec{Twisting and Klein factors.}

It is well known that statistics of fields can be modified by
so-called Klein factors. If the original field operators
$\Phi_j(x_j)$ satisfy the braiding relation:
\eqn\IFib{ \Phi_1(x_1)\Phi_2(x_2) = {\CR}_{12} \Phi_2(x_2) \Phi_1(x_1)
\qquad ;\qquad x_1<x_2}
then the fields $\tilde \Phi_j(x_j)$ modified by the Klein factors
satisfy the following braiding relation:
\eqn\IFiib{ \tilde \Phi_1(x_1)\tilde \Phi_2(x_2)
= \tilde \CR_{12} \tilde \Phi_2(x_2) \tilde \Phi_1(x_1)
\qquad ;\qquad x_1<x_2}
with
\eqn\IFiiib{ \tilde \CR_{12} = F_{12}\ \CR_{12}\ F^{-1}_{21} }
for some $F$. This modification
corresponds to a gauge transformation of the braiding matrix $R_{12}$.
The new braiding matrix is still a solution of the Yang-Baxter
equation provided that the factor $F$ satisfies a 2-cocycle relation:
\eqn\IFivb{
(1\ot\De)(F)(1\ot F)\ =\ (\De\ot 1)(F)(F\ot 1) }
Since, as we explained in the previous sections, the comultiplication
of the quantum symmetry algebra is encoded in the braiding relations,
twisting the braidings will imply a twisting of the comultiplication.
The new comultiplication $\tilde \De$ is given by:
\eqn\IFvb{ \tilde \De(a)\ =\ F^{-1}\De(a) F \qquad;\qquad a\in\CA}
It  is still coassociative thanks to the cocycle condition \IFivb~.
This twisting is a particular example of the Drinfel'd twist for quasi-Hopf
algebras \ref\rtwist{V. G. Drinfel'd, Leningrad Math. J. 1 (1990)
1419}.

The simplest construction of Klein factors
is provided by the cocycle operators involved
in the vertex operator representation of current algebras.
This amounts to multiplying  the fields by operators with appropriate
commutation relations which are encoded into a 2-cocycle.

\subsec{Remarks}

We conclude this section by emphasizing that, as we have developed them,
the algebras $\CA$ and $\adu$ are on completely different footing.
The elements of $\CA$ are integrals of conserved currents, and the action
of $e_a$ on a field is a generic adjoint action.  This is not true for
elements of $\adu$.  This fact is quite clear in the current
representation, when one considers the integrated form of
\IEv.  This is not an adjoint action since
\eqn\nooot{
e^i (Q_a) = \mu^{ib}_a (Q_b) \not=
ad_{e^i} \( Q_a \) \equiv m^i_{cd} \> e^c \> Q_a \> s(e^d ). }

This asymmetry between the algebras $\CA$ and $\adu$ is quite similar in
spirit to what occurs for the finite quantum group symmetry of minimal
conformal
field theory\rgs\rfw.
There, one Borel subalgebra $\CB_+$ of $SL(2)_q$ is realized as screening
operators, and are thus integrals of conserved currents.  The other half
of $SL(2)_q$ arises as `contour annihilation' operators, which
are defined operationally via the conformal transformations of the field
multiplets.
Thus in  the conformal field theory context $\CA = \CB_+$ and
the full $SL(2)_q$ algebra arises in the double $\CD (\CA )$.
In this paper
the action of $e^a \in \CA^*$ is defined operationally
via the finite rotations
$R_{\pm \pi}$ rather than  conformal transformations, as the latter
are not generally symmetries of the
models.  We remark however that for the affine $\hat{SL(2)}_q$
symmetry of the sine-Gordon theory, the  complete $\hat{SL(2)}_q$
algebra is realized in terms of conserved currents and thus constitutes
that algebra $\CA$, which is quite different than the situation in
conformal field theory; in this model one must actually deal with
$\CD (\hat{SL(2)}_q )$.

\def\y{\CY}
\def\dy{{ \CD (\CY) }}
\def\f#1#2#3{{ f^{#1#2#3} }}
\def\q#1#2{{ Q_{#1}^{#2} }}

\subsec{Yangian Symmetry}

In this subsection we illustrate the above construction for the
case of Yangian symmetry.  As described in the introduction, integrable
quantum field theories which are invariant under some finite Lie group
$G$ generically have a larger symmetry corresponding to the Yangian $\y$.

Let $J_\mu^a (x), a=1,..,{\rm dim}(G)$ denote the local currents for the
symmetry $G$, and $\q 0a$ the conserved charges which generate the Lie
algebra of $G$~~\foot{For $G= SL(2)$ we take $\f abc = i \epsilon^{abc}$.}:
\eqn\IFi{
[\q 0a , \q 0b ] = \f abc \q 0c . }
The non-local currents which generate the additional Yangian conserved
charges are given by \lus\yang
\eqn\IFii{
\CJ^a_\mu (x) = z(\delta) \epsilon_{\mu\nu} J^a_\nu (x)
-\al \f abc J^b_\mu (x) \int_{-\infty}^{x-\delta}
J^c_t (y) dy . }
In this expression, $z(\delta )$ is a constant which serves to
remove the singularity at $\delta = 0$.  The current $\CJ_\mu^a$
is normalized such that the constant $\al$ is
\eqn\IFal{
\al = - \frac{2\pi i}{c_A} , }
where $c_A$ is the Casimir
in the adjoint representation: $\f abc \f bcd = -c_A \delta^{ad}$.
($c_A = 2$ for SL(2).)
Let $\q 1a$ denote the conserved charge for the non-local current,
i.e. $\q 1a = \int_{-\infty}^{\infty} dx \CJ_t^a (x)$.
The charge $\q 1a$ is $G$-covariant:
\eqn\IFiii{
[\q 0a , \q 1b ] = \f abc \q 1c . }

One may easily apply the general construction in
sections 2.1-2.3.  The result is
\eqn\IFiv{\eqalign{
\De (\q 0a ) &= \q 0a \ot 1 + 1 \ot \q 0a \cr
\De (\q 1a ) &= \q 1a \ot 1 + 1 \ot \q 1a
+ \al \> \f abc \q 0b \ot \q 0c \cr
\vep (\q 0a ) &= \vep (\q 1a ) = 0\cr
s( \q 0a ) &= - \q 0a ;~~~~~s(\q 1a ) = -\q 1a + \al \f abc \q 0b \q 0c .
\cr }}
The adjoint actions are
\eqn\IFv{\eqalign{
ad_{\q 0a} (\Phi ) &= \[ \q 0a , \Phi \] \cr
ad_{\q 1a} (\Phi ) &= \[ \q 1a , \Phi \] - \al \f abc [\q 0b , \Phi ] \q 0c .
\cr }}

The equations \IFi, \IFiii, and \IFiv\ are the defining relations of
the Yangian $\y$, which plays the role of the Hopf algebra
$\CA$  above. The Yangian, as defined by Drinfel'd\rdrin, is a deformation
of the positive half of the affine Lie algebra, with deformation
parameter $\al$ ($\al =0$ is the undeformed case).  Though the
relations \IFi, \IFiii\ are undeformed, the deformation \IFiv\ of the
comultiplication implies that other relations of the affine Lie algebra
must be deformed.  Consider for example the case of SL(2).  Define
$\q 2a = -\f abc \q 1b  \q 1c $. (For SL(2) this definition
is equivalent to $[\q 1a , \q 1b ] = \f abc \q 2c$.)  Then one has
deformed Serre relations\rdrin\foot{The following specific form of
the deformed Serre relations is only valid for the SL(2) Yangian.}
\eqn\IFvi{
[\q 2a , \q 1b ] + [\q 2b , \q 1a ]
= \al^2
\( \f acd \{ \q 0b , \q 0c , \q 1d \}
+ \f bcd \{ \q 0a , \q 0c , \q 1d \} \) , }
where $\{~ \}$ denotes the symmetrized product:
\eqn\symm{
\{ x_1 , x_2 , ..., x_n \} = \inv{n!}
\sum_{i_1 \neq i_2 \neq \cdots \neq i_n }
x_{i_1} x_{i_2} \cdots x_{i_n} . }
Obviously, for the usual affine Lie algebra the left hand side of
\IFvi\ is zero.

We define $\yd$ as the dual algebra to $\y$, and $\dy$ as the
Yangian double.  $\dy$ is generated by elements $\q {-1}a$.
The algebraic relations in $\dy$ can be determined from \IDvi,
\IDvii\ with $(\q 0a )^* \equiv 2\al \q {-1}a $, as was done in
\rls. One thereby understands $\dy$ as a deformation of the full
affine Lie algebra.  We will give a different presentation of
$\dy$ in the subsequent sections.  For now, we illustrate the
geometrical meaning of the quantum double for the current multiplet.
  From \IEv\  and \IFiv\ one obtains
\eqn\IFvii{
\q {-1}a \( \CJ_\mu^b (x) \) = -  \frac{1}{2} \f abc \> J^c_\mu (x) . }
Thus one sees that $(\q 0a )^*$ acting on $\CJ_\mu^b (x)$ essentially
removes the string, leaving $-\al \f abc J^c_\mu (x)$.

\vfill\eject

\newsec{Dynamical Properties}

In this section we describe how the affine quantum group symmetry
fixes the S-matrix.  We also discuss some aspects of the form factors.
As stated in the introduction, the basic example we develop is the
SL(2) Yangian.

\subsec{The S-matrix from the Universal $\CR$-matrix}

The spectrum of the basic SL(2) Yangian invariant model consists of
a doublet of solitons, transforming in the 2-dimensional  representation of
SL(2).  Formulated as a current-current perturbation of the WZW model
\i, these solitons are created by the only chiral (or antichiral) primary
fields of the WZW theory, which are  in the 2-dimensional representation
and have Lorentz spin $\pm 1/4$\rkz.
Denote the single particle states as $\ket \th \in V$, where $V \simeq
\Cmath^2$ is
the 2-dimensional  vector space, and $\th$ is the rapidity:
\eqn\IIIi{
E= m\cosh (\th ); ~~~~~P= m \sinh (\th ). }
The representation of the charges $\q 0a$ on the states is taken to be
\eqn\IIIib{
\rho_V (\q 0a ) \ket \th = s^a \ket \th , }
where $s^a = \sigma^a /2$.  ($\sigma^a$ are the Pauli spin matrices.)

The representation of the charges $\q 1a$ on the states can be
deduced as follows. It was shown in \rls\ that one has the
following fundamental relation between the energy momentum
tensor $T_{\mu\nu} (x)$ and the local currents $J_\mu^a (x)$:
\eqn\IIIii{
\[ \q 1a , T_{\mu\nu} (x) \]
= -\inv{2} \( \ep_{\mu\al} \d_\al J^a_\nu (x) + \ep_{\nu\al}
\d_\al J^a_\mu (x) \). }
The above relation was proven by comparing the form factors of
$T_{\mu\nu}$ and $J^a_\mu$, and using the Ward identities for
the affine quantum group symmetry.  The relation \IIIii\ can
also be established by showing that $[\q 1a, T_{\mu\nu} (x)]$ is
local with respect to $J^a_\mu$; Lorentz  and  SL(2) covariance
then fixes the right hand side of \IIIii.
Let $L$ be the generator of Lorentz boost:
$$L= - \int dx \> x T_{tt} (x)~~~~~~~~~~~~~(t=0).
$$
Then integration of \IIIii\ yields
\eqn\IIIiii{
[L , \q 1a ] = -\q 0a . }
On shell, a finite Lorentz boost of rapidity $\beta$ shifts
$\th \to \th -\beta$.  Therefore, on shell, $L= \CL \equiv
-\der{\th}$.  One thus deduces that
\eqn\IIIiv{
\rho_V^\th (\q 1a ) \ket {\th} = \th \> s^a \> \ket {\th} . }

More generally, the role of the Lorentz boost operator in the
theory of affine quantum groups can be formulated algebraically
as follows.  Define $t_\th$ to be an automorphism of
$\da$ such that $\rho_V^\th (a) = \rho_V \( t_\th (a) \)$,
$a\in \da$.  In the case of the Yangian this automorphism takes the
form\rdrin
\eqn\IIIv{
t_\th (\q 1a ) = \q 1a + \th \q 0a ,~~~~~t_\th (\q 0a ) = \q 0a . }
The automorphism $t_\th$ corresponds to a finite Lorentz boost
of rapidity $\th$: $t_\th (P_\pm ) = e^{\mp \th} P_\pm$,
where $P_\pm$ are the light-cone components of the momentum.
A Lorentz boost of rapidity $\th$ is equivalent to a Euclidean
rotation by the same angle.  In the algebra $\CA$, one has
\eqn\IIIvi{
\[ L, a \] = \CL \( t_\th (a) \). }

The fact that the non-local charges $\q 1a$ act on the states in this
momentum dependent fashion is the primary reason why they provide
strong dynamical constraints on the theory.  We now construct the
S-matrix using only the quantum symmetry by relating it to the
universal $\CR$-matrix.   The two particle to two particle S-matrix is
an operator
\eqn\IIIvii{
S_{12} (\th_1 - \th_2 ) :  ~~~~V_1 \ot V_2 \to V_2 \ot V_1 . }
It is required to commute with the quantum symmetry:
\eqn\IIIviii{
S_{12} (\th_1 - \th_2 ) ~
\rho_{V_1}^{\th_1} \ot \rho_{V_2}^{\th_2}
\( \Dep (a) \)
=
\rho_{V_1}^{\th_1} \ot \rho_{V_2}^{\th_2}
\( \De (a) \) ~
S_{12} (\th_1 - \th_2 ) , ~~~~~a\in \CA. }
The above symmetry equation is enough to fix the S-matrix up to an overall
scalar function of $\th_{12} \equiv \th_1 -\th_2$, and the solution
automatically
satisfies the Yang-Baxter equation. This follows from well-known
results in the theory of affine quantum groups\rdrin\rjim.  In the case of the
Yangian it fixes $S \propto (\th + \al P )$,
where $P$ is the permutation operator ($P v_1 \ot v_2 = v_2 \ot v_1$).
Since the dual algebra $\adu$ has so far played no role in the equation
\IIIviii, it is natural to suppose that the full quantum double structure
fixes in addition the overall scalar factor, as was argued by
Smirnov\rsm.

Given the representation $\rho_V^\th$ of $\CA$, and the relations \IDvi,
one can in principle deduce the representation $\rho_V^\th$ in the
full quantum double $\da$.  Consider then evaluating the universal
$\CR$-matrix in this representation:
\eqn\IIIix{
R^{V_1 V_2} (\th_1 , \th_2 ) = \sum_a \rho_{V_1}^{\th_1} (e_a ) \ot
\rho_{V_2}^{\th_2} (e^a ) . }
The matrix $R^{V_1 V_2}$ satisfies several important identities
as a consequence of the quantum double structure.
Where there is no chance for confusion, we will simplify the notation
and write $R_{12} =R^{V_1 V_2}$.
Due to the automorphism
$t_\th$, $R_{12}$ is only a function of $\th_{12}$.  By construction it
satisfies the $\CA$-symmetry equations \IIIviii.  It also automatically
satisfies the rapidity-dependent Yang-Baxter equation, as a consequence
of (A.9):
\eqn\IIIx{
R_{12} (\th_{12}) R_{13} (\th_{13} ) R_{23} (\th_{23} )
= R_{23} (\th_{23} ) R_{13} (\th_{13} ) R_{12} (\th_{12} ).}

There are two additional relations satisfied by $R_{12} (\th )$ which
are related to crossing symmetry and unitarity.  The unitarity
condition reads
\eqn\IIIxi{
R_{12} (\th ) R_{21} (-\th ) = 1.}
(In the above equation $R_{21} = P R_{12} P$.) A proof of \IIIxi\ may
be given along the following lines. Since $R$ satisfies the
equation (A.6), then $R^\prime \equiv PRP$ satisfies
$R^\prime \De = \De^\prime R$.  Since the equation (A.6)  determines
$R$ up to an overall scalar factor and assuming that the tensor product
$V_1\otimes V_2$ is irreducible for generic values of $\th$,
$R^\prime$ must be proportional
to the inverse of $R$:  $R_{12} (\th ) R_{21} (-\th ) = f(\th )$,
where $f(\th )$ is some scalar function.  Now using the quasi-triangularity
property (A.11), one finds
$(\De \ot 1 ) R_{12} R_{21} = R_{23} R_{13} R_{31} R_{32}$.  The latter
implies $f(\th )= (f(\th ))^2$, or $f(\th ) = 1$.

The crossing symmetry property may be derived as follows.  Suppose that in
a particular representation of $\CA$ the following formula is valid
\eqn\IIIxii{
\rho_V^\th \( s' (a) \) = C \( \rho_V^{\th + i\pi } (a) \)^t C^{-1}, }
where
$C$ is a constant `charge  conjugation' matrix, and $t$ denotes transpose.
Indeed it is clear from \IBi, and from the equivalence of Lorentz boosts
and Euclidean rotations, that the antipode $s'$ should shift $\th$ by
$i\pi$.  The transpose $t$ ensures $s' (ab) = s'(b) s'(a)$.
Then (A.14) implies
\eqn\IIIxiii{
R^{-1}_{12} (\th ) = C_1 R_{12}^{t_1} (\th + i\pi ) C_1^{-1} , }
thus
\eqn\IIIxiv{
\( R_{12} (\th ) \) \( C_1 R_{12}^{t_1} (\th + i\pi ) C_1^{-1} \)
=1. }
Combining \IIIxi\ and \IIIxiv\ on obtains the crossing relation
\eqn\IIIxv{
R_{21} (\th ) = C_1 R_{12}^{t_1} (i\pi - \th )C_1^{-1} . }

Let us now specialize to the Yangian.  One finds
$\rho_V^\th \( s^\prime (a) \) = C \( \rho_V^{\th - \al} (a) \)^t C^{-1}$,
with
$$C=\left( \matrix{ 0&1\cr -1&0\cr} \right) .$$
(Recall $\al = -i\pi$.)   One may express $R_{12}$ as
\eqn\IIIxvi{
R_{12} (\th ) = v(\th ) (\th + \al P ), }
where
$v(\th)$ is a scalar function\rdrin.  Then \IIIxi\ and \IIIxiv\ imply
the following functional relations for $v(\th )$:
\eqna\IIIxvii
$$\eqalignno{
v(\th ) v (-\th ) &= \inv{ \al^2 - \th^2 }  &\IIIxvii {a}\cr
v(\th ) v(\th - \al ) &= \inv{\th^2 - \al^2}  &\IIIxvii {b}\cr
\Rightarrow ~~~v(\th ) &= -v(-\th -\al ) . &\IIIxvii {c}\cr }$$
Related arguments for fixing overall scalar factors were given
in \rdrin\rfr\rsm.

\def\tt{{ \tilde{T} }}

The formula \IIIxvii{b}\ is equivalent to requiring that the
quantum determinant is $1$.
To see this, let $\tt (\th) \equiv \rho_V^\th (e_a) \ot e^a$
denote a matrix with elements $\tt_{ij} \in \adu$, $ i,j  \in \{ 1,2 \}$.
Then the quantum determinant of $\tt$ is defined as
\ref\rks{P. P. Kulish and E. K. Sklyanin, J. Soviet Math. 19, p61.}
\eqn\IIIxviii{\eqalign{
Det_q \( \tt (\th ) \) \equiv
\inv{2} ( \tt_{11} (\th ) & \tt_{22} (\th + \al )
+ \tt_{22} (\th ) \tt_{11} (\th + \al ) \cr
& - \tt_{12} (\th )
\tt_{21} (\th + \al ) - \tt_{21} (\th ) \tt_{12} (\th + \al )
) . \cr}}
Using the fact that $\rho_V (\tt ) = R$, one can show that
$\rho_V (Det_q (\tt ) ) = 1$ implies \IIIxvii{b}.

\def\dt{{ \tilde{\De} }}

In order to compare with the conventional physical S-matrix, one must
consider a twisted $R$-matrix along the lines developed in section 2.6.
This is due to the fact that the soliton fields which create the particles
have Lorentz spin $\pm 1/4$, and therefore have non-trivial braiding among
themselves and with the local currents $J_a^\mu  (x)$.
Let $\{ \q 03 , \q 0\pm \}$ denote a basis for the charges $\q 0a$, with
$[\q0 3 , \q 0\pm ] = \pm 2 \q 0\pm$, $[\q 0+ , \q 0- ] = \q 03$, and
similarly for $\q 1a$.  Let $J^\pm_\mu (x) , J^3_\mu (x)$ denote
the corresponding currents, and $\Psi_V (x)$ the soliton fields.  It
is clear from a bosonized description (see e.g. \nlc\ and references
therein) that one has
\eqn\IIIxix{\eqalign{
J^\pm_\mu (x) \> \Psi_V (y) &= (-1)^{\pm \hat{Q}_0^3 } \( \Psi_V (y) \)
\> J^\pm_\mu (x) ~~~~~x>y   \cr
J^3_\mu (x) \> \Psi_V (y) &= \Psi_V (y) \> J^3_\mu (x) . \cr } }
In accordance with the general principles outlined in section 2,
in determining the S-matrix one should take the twisted comultiplication
$\tilde{\De}$:
\eqn\IIIxx{\eqalign{
\dt (\q 03 ) &= \q 03 \ot 1 + 1 \ot \q 03 \cr
\dt (\q 0\pm ) &= \q 0\pm \ot 1 + (-1)^{\pm \q 03} \ot \q 0\pm \cr
\dt (\q 1\pm ) &= \q 1\pm \ot 1 + (-1)^{\pm \q 03} \ot \q 1\pm
\mp \inv{2} \al
\( (-1)^{\pm \q 03} \q 03 \ot \q 0\pm - \q 0\pm \ot \q 03 \)
\cr
\dt(\q 13 ) &= \q 13 \ot 1 + 1\ \ot \q 13
+ \al
\( (-1)^{\q 03} \q 0- \ot \q 0+
- (-1)^{-\q0 3} \q 0+ \ot \q 0- \) . \cr }}
One may check that this twisted comultiplication is still a
homomorphism of the Yangian relations \IFi\ and \IFiii, i.e.
(A.1d) holds with $\De \to \dt$.
This twisted comultiplication may also be understood as a
consequence of the fact that the Yangian invariant model
arises as the $q\to -1$ limit of the $\hat{SL(2)}_q$ invariant
sine-Gordon model\nlc.

The above twist of the comultiplication implies
(via \Iaxi\ with $\De \to \dt$) that the antipodes are modified to
\eqn\IIIxxi{\eqalign{
s(\q 03 ) &= -\q 03 , ~~~~~ s( \q 0\pm ) = - (-1)^{\mp \q 03} \q 0\pm
\cr
s( \q 13 ) &= -\q 13 - \al \q 03 ,~~~~~
s( \q 1\pm ) = (-1)^{\mp \q 03} \( - \q 1\pm - \al \q 0\pm \) . \cr}}
Consequently, one finds that \IIIxii\ is valid, but with
$C\to \tilde{C}$,
\eqn\IIIxxii{
\tilde{C} = \left( \matrix{0&1\cr 1&0\cr } \right) . }

\def\vt{{ \tilde{v} }}
\def\tv{{ \tilde{v} }}

Repeating the above analysis with the twisted $R$-matrix
$\tilde{R}$, one finds it is still crossing symmetric and
unitary.  $\tilde{R}_{12} $ may be expressed as
\eqn\IIIxxiii{
\tilde{R}_{12} = \tv (\th )
\left( \matrix{ -\al -\th &0&0&0\cr
0&\th&-\al&0\cr
0& -\al & \th & 0 \cr
0& 0 & 0 & -\al -\th \cr} \right) . }
The functional relations for $\tv$ are now changed to
\eqn\IIIxxiv{
\tv (\th ) \tv (-\th ) = 1 , ~~~~~~\tv (\th ) = \tv (-\al -\th ) . }

The functional relations \IIIxxiv, together with some knowledge of
the asymptotic development of $\tilde{R} (\th )$ as $\th \to \pm \infty$,
are enough to determine a solution.  Expressing
$\tilde{R}_{12} = \rho_V (e_a ) \ot \rho_V^{-\th } (e^a )$, and noting
that $\rho_V^{-\th } (e^a ) $
only has zero or negative powers of $\th$, one sees
that $\tv (\th ) =  \Cmath [[ \th^{-1} ]] /\th $.  Thus
\eqn\IIIxxv{
\vt (\th \to \pm \infty ) \to \pm \frac{i}{\th } . }
In the limit $\th \to \pm \infty$, $\tilde{R} (\th )$ becomes the
braiding matrix of the solitons:
\eqn\IIIxxvi{
\tilde{R}_{12} (\th \to \pm \infty ) \to
\pm {\rm diag} ( -i, i, i, -i ) . }
A solution to the functional equations \IIIxxiv, which has the proper
asymptotic development \IIIxxv, may be constructed iteratively to yield
an infinite product expression:
\eqn\IIIxxvii{
\tv (\th ) = \inv{\th - \al }
\prod_{j=1}^\infty
\[
\( \frac{ -\al (2j) + \th }{ -\al (2j) - \th } \)
\( \frac{-\al (2j+1) - \th }{-\al (2j+1) + \th } \)
\] . }
Using standard identities this may be expressed as
\eqn\IIIxxviii{
\tv (\th ) = \inv{\th + \al } \>
\frac{ \Gamma (1/2 - \th /2\al ) \Gamma (\th /2\al ) }{\Gamma (1/2 + \th /2\al
)\Gamma ( -\th /2\al )} . }
One may check using Sterling's formula that $\tv (\th )$ has the
asymptotics \IIIxxv.   Comparing with the known physical S-matrix,
we conclude that this S-matrix is precisely characterized by the
twisted universal $\CR$-matrix:
\eqn\IIIxxix{
S_{12} (\th ) = \tilde{R}_{12} (\th ) .   }

\subsec{Multiplicative Presentations of the Quantum Double from
Vertex Operators}

Let us formally introduce generalized creation-annihilation
operators $Z(\th )$, valued in the particle vector space $V$,
such that the multiparticle asymptotic states correspond to the
action of $Z(\th )$'s on the vacuum:
\eqn\IIIBo{
\ket{\th_1 , ..., \th_n } = Z(\th_1 ) \cdots Z(\th_n ) \ket {0} . }
Though we have only formally introduced the operators $Z(\th )$, they
may be given the following interpretation.  Consider a free scalar field
theory.  Normally this field may be expanded into creation-annihilation
operators $a(\th ), a^\dagger (\th )$.  This may be written as
\eqn\IIIBi{
\phi (x) = \oint_\CC d\th \> Z(\th ) e^{-i p(\th ) \cdot x } , }
where the contour $\CC$ in the $\th$-plane runs along the real axis and
returns along the line $\th = i\pi$.  Namely, $Z(\th ) = a(\th )$ and $
Z(\th + i\pi ) = a^\dagger (\th )$ for $\th$ real.  The
$Z(\th )$ operators may roughly be thought of as a generalization
of this to an interacting theory.

The operators $Z(\th )$ are defined to be characterized by their
intertwining relations:
\eqn\IIIBii{\eqalign{
ad_{e_a} \( Z(\th ) \) &= \rho_V^\th (e_a ) \; Z(\th ) \cr
e^a  \( Z(\th ) \) &= \rho_V^\th (e^a ) \; Z(\th ) . \cr }}
They are thus by definition vertex operators for the affine quantum group.
Such vertex operators were defined and studied in \rfr.
They are also defined to have S-matrix exchange relations:
\eqn\IIIBiii{
Z_1 (\th_1 ) Z_2 (\th_2 ) = S_{12} (\th_{12} ) Z_2 (\th_2 ) Z_1 (\th_1 ), }
where the subscripts refer to the vector spaces $V_1 , V_2$.  For
simplicity we here ignore complications due to Klein factors and take
$S_{12} (\th ) = R_{12} (\th )$.  By applying $ad_{e_a}$ to both sides
of \IIIBiii\ one derives that $S_{12}$ must satisfy the defining
relations of the universal $\CR$-matrix, so that
\IIIBiii\ may be viewed as a consequence of \IIIBii.

The vacuum is defined to satisfy
\eqn\IIIBiv{\eqalign{
e_a \ket{0} &= \ep_a \ket{0} , ~~~~~e^a \vac = \ep^a \vac \cr
\lvac e_a &= \lvac \ep_a ,~~~~~\lvac e^a = \lvac \ep^a , \cr}}
which is the statement that the vacuum is `annihilated' by
the elements of $\da$.  It is evident from the properties in the
appendix that the counits form one-dimensional representations of
$\da$:  $\ep_a \ep_b = m_{ab}^c \ep_c ,\  \ep^a \ep^b = \mu^{ba}_c \ep^c$.
One also has
\eqn\IIIBv{\eqalign{
e_a ~ Z(\th_1 ) \cdots Z(\th_n )
&= \mu^{bc}_a ad_{e_b} \( Z(\th_1 ) \cdots Z(\th_n ) \) e_c \cr
e^a ~ Z(\th_1 ) \cdots Z(\th_n )
&= m^a_{bc}  e^b \( Z(\th_1 ) \cdots Z(\th_n ) \) e^c \cr
 Z(\th_1 ) \cdots Z(\th_n )~ s(e_a )
&= \mu^{bc}_a  s(e_b ) ad_{e_c} \( Z(\th_1 ) \cdots Z(\th_n ) \)  \cr
 Z(\th_1 ) \cdots Z(\th_n )~ s(e^a )
&= m^a_{bc}  s(e^b ) e^c \( Z(\th_1 ) \cdots Z(\th_n ) \)  . \cr
}}
Combined with \IIIBiv, and the analog of \ICiii,
the above relations imply for instance
\eqn\IIIBvi{\eqalign{
e_a ~ Z(\th_1 ) \cdots Z(\th_n ) \ket{0}
&= ad_{e_a} \( Z(\th_1 ) \cdots Z(\th_n )
\) \ket{0} \cr
&=
\rho_{V_1 \cdots V_n} (e_a ) Z(\th_1 ) \cdots Z(\th_n  ) \ket{0} , \cr }}
where
\eqn\dodo{
\rho_{V_1 \cdot\cdot V_n} (e_a ) =
\mu^{b_1 c_1}_a \, \mu^{b_2 c_2}_{c_1} \cdots \mu^{b_{n-1} b_n}_{c_{n-2}}
\rho_{V_1} (e_{b_1} ) \ot \cdots \ot \rho_{V_n} (e_{b_n} ) . }

It was understood by Smirnov that the elements of $\da$ can be reconstructed
from the algebra of vertex operators.  Here, we clarify some aspects of
this idea.  Define a matrix $L(\th )$ of operators, with matrix elements
\eqn\IIIBvii{
{L_a}^b (\th ) =
2\pi i \res_{\th' = \th -i\pi } ~
Z_a (\th ) Z_c (\th '  ) C^{cb} . }
It is not difficult to prove using only the exchange relation \IIIBiii\
and crossing symmetry \IIIxv\ that $L$ satisfies the following
relation
\eqn\IIIBviii{
S_{21} (\th_{21} ) L_1 (\th_1 ) S_{12} (\th_{12} ) L_2 (\th_2 )
= L_2 (\th_2 ) S_{21} (\th_{21} ) L_1 (\th_1 ) S_{12} (\th_{12} ) , }
where
$L_1 = L\ot 1, L_2 = 1\ot L$.

Relations of the kind \IIIBviii\ are well-known in the theory of quantum
groups and are referred to as multiplicative presentations\rdrin\rRS.
Define two matrices of operators $T^\pm (\th )$ as follows
\eqn\IIIBix{\eqalign{
T^+ (\th ) &= \sum_a  e_a \rho_V^\th (e^a ) \cr
T^- (\th ) &= \sum_a \rho_V^\th ( s'(e_a )) e^a . \cr }}
The matrix elements of $T^+$ are in $\CA$, whereas those of $T^-$ are
in $\adu$.  From the Yang-Baxter equation (A.9)  evaluated in the
appropriate representations, one obtains
\eqn\IIIBx{\eqalign{
R_{12} (\th_{12} ) T_2^\pm (\th_2 ) T_1^\pm (\th_1 )
&= T_1^\pm (\th_1 ) T^\pm_2 (\th_2 ) R_{12} (\th_{12} ) \cr
R_{12} (\th_{12} ) T_2^+ (\th_2 ) T_1^- (\th_1 )
&= T_1^- (\th_1 ) T_2^+  (\th_2 ) R_{12} (\th_{12} ) . \cr}}
One can also show using \IIIBix\ and the basic properties of $\da$ that
\eqn\IIIBxi{
T^\pm (\th ) s' (T^\pm  (\th ) ) = s' (T^\pm  (\th ) ) T^\pm  (\th ) = 1, }
and
\eqn\IIIBxii{
\Dep \( T^\pm (\th ) \) = T^\pm (\th ) \ot T^\pm (\th ) . }
The algebra defined by \IIIBx-\IIIBxii\ is isomorphic to $\da$,
provided one also imposes that the quantum determinant of $T^\pm$
is $1$.  Finally define
\eqn\IIIBxiii{
T(\th ) \equiv T^+ (\th ) s' \( T^- (\th ) \) . }
Then using the relations \IIIBx\ one can show
\eqn\IIIBxiv{
R_{21} (\th_{21} ) T_1 (\th_1 ) R_{12} (\th_{12} ) T_2 (\th_2 )
= T_2 (\th_2 ) R_{21} (\th_{21} ) T_1 (\th_1 ) R_{12} (\th_{12} ) . }

If $T$ satisfies \IIIBxiv, then it is not difficult to show
using the unitary property of $R_{12}$, that
$T-1$ does also.  The consistent identification is
\eqn\IIIBxivb{
L = T-1 . }
This conclusion is reached by studying the counits.  By construction
$\ep (T) = 1$.  On the other hand, from the previously given
definition of the counit one finds
\eqn\IIIBxv{
\ep (L_a^b (\th ) ) =
2\pi i  \res_{\th' = \th -i\pi }
\lvac Z_a (\th ) Z_c (\th' ) C^{cb} \vac = 0. }
Note that the vertex operator algebra only yields the relation
\IIIBviii, thus one must assume that $L+1$ can be factorized as in
\IIIBxiii\ in order to extract the $\da$ relations \IIIBx.

The intertwining property \IIIBii~ of the operators $Z(\th)$ can also be
derived
from the exchange relation \IIIBiii~ and the reconstruction formula \IIIBvii~
of the operators $L(\th)$. A simple computation gives~:
\eqn\IIIBcov{
L_1(\th_1)S_{12}(\th_{12})Z_2(\th_2) =
S_{12}(\th_{12}) Z_2(\th_2) L_1(\th_1) }
and similarly with $T=L+1$. Eq. \IIIBcov~ is the multiplicative form
of the intertwining relations \IIIBii . Notice that the set of relations
\IIIBiii , \IIIBviii~ and \IIIBcov~ is very reminiscent of the defining
relation of the quantum cotangent bundle
\ref\AlFa{A.Yu. Alekseev and L.D. Faddeev, Commun. Math. Phys. 141 (1991) 413}.
Finally, one can also show from \IIIBix\ and \IIIBvi\ that
\eqn\IIIBxvb{\eqalign{
T^+_0 (\th_0 ) ~
Z_1 (\th_1 ) \cdots Z_n (\th_n ) \vac
&=
S_{10} (\th_{10} ) S_{20} (\th_{20} )\cdots S_{n0} (\th_{n0} )
Z_1 (\th_1 ) \cdots Z_n (\th_n ) \vac
\cr
T^-_0 (\th_0 ) ~
Z_1 (\th_1 ) \cdots Z_n (\th_n ) \vac
&=
S_{10} (\th_{10} ) S_{20} (\th_{20} )\cdots S_{n0} (\th_{n0} )
Z_1 (\th_1 ) \cdots Z_n (\th_n ) \vac
. \cr }}

\medskip
\subsec{Multiplicative Presentations of the Yangian Double}

In \rls\ the Yangian Double was studied using directly the
formulas in appendix A.  In this section we develop more fully
the multiplicative presentation of the last subsection.
Though it is evident from \IIIBix\ that one can extract the
$\da$ algebra from \IIIBx, this exercise is useful in providing
a concrete basis for the generators where the relations are finite,
and will make it evident that $\dy$ is a deformation of the full
affine Lie algebra.

\def\t#1#2#3{{ t^{(#1)}_{#2#3} }}

Expand $T^\pm (\th )$ as power series in $\th$ as follows:
\eqn\IIIBxvi{\eqalign{
T^+ (\th ) &= 1 - 2\al \sum_{n=0}^\infty t^{(n)} \th^{-n-1} \cr
T^- (\th ) &= 1 + 2\al \sum_{n=0}^\infty t^{(-n-1)} \th^{n} .  \cr}}
In the relations \IIIBx\ one can cancel the factor $v(\th )$ in
\IIIxvi\ from both sides of the equation to obtain
\eqn\doggle{\eqalign{
\[ T_{ij}^\pm (\th_2 ) , T_{kl}^\pm (\th_1 ) \]
& = \frac{\al}{\th_1 - \th_2 }
\( T_{kj}^\pm (\th_1 ) T^\pm_{il} (\th_2 ) - T_{kj}^\pm (\th_2 )
T_{il}^\pm (\th_1 ) \)  \cr
\[ T_{ij}^+ (\th_2 ) , T_{kl}^- (\th_1 ) \]
& = \frac{\al}{\th_1 - \th_2 }
\( T_{kj}^- (\th_1 ) T^+_{il} (\th_2 ) - T_{kj}^+ (\th_2 )
T_{il}^- (\th_1 ) \) .  \cr }}
Let $\t nij , i,j = 1,2$
denote the matrix elements of $t^{(n)}$.  Then the above formulas
imply
\eqn\IIIBxvii{\eqalign{
\[\t nij , \t mkl \]
&= \inv{2} \( \de_{il} \t {m+n}kj - \de_{kj} \t {m+n}il \)
+ \al \sum_{p\leq m-1}
\( \t {m-p-1}kj \t {n+p}il - \t {n+p}kj \t {m-p-1}il \) \cr
\[\t {-n}ij , \t {-m}kl \]
&= \inv{2} \( \de_{il} \t {-m-n}kj - \de_{kj} \t {-m-n}il \) \cr
&~~~~~~~~~~+ \al \sum_{p\leq n-1}
\( \t {-m-p-1}kj \t {p-n}il - \t {p-n}kj \t {-m-p-1}il \) \cr
\[\t {n}ij , \t {-m}kl \]
&= \inv{2} \( \de_{il} \t {n-m}kj - \de_{kj} \t {n-m}il \) \cr
&~~~~~~~~~~+ \al \sum_{p\leq min(m-1,n-1)}
\( \t {p-m}kj \t {n-p-1}il - \t {n-p-1}kj \t {p-m}il \) , \cr}}
\eqn\IIIBxviii{\eqalign{
\Dep \( \t nij \) &=
\t nij \ot 1 + 1\ot \t nij - 2\al \sum_{p\leq n-1} \t pik \ot \t {n-1-p}kj
\cr
\Dep \( \t {-n}ij \) &=
\t {-n}ij \ot 1 + 1\ot \t {-n}ij +
2\al \sum_{p\leq n-1} \t {-p-1}ik \ot \t {-n+p}kj
. \cr}}
(Above, $\t nij$ ($\t {-n}ij$) has $n\geq 0$ ($n\geq 1$).)
In deriving these relations we used the identity
$1/(\th -\th ') = (1+\sum_{n=1}^\infty (\th' /\th )^n )/ \th $.

It is clear that when $\al = 0$, the algebra $\dy$ defined in \IIIBxvii\ is
isomorphic to the level 0 SL(2) affine Lie algebra.  The isomorphism is
$t^{(n)} = \sum_{a=1}^3  Q_n^a s^a$, $n\in \Z$, with
$[ \q na , \q mb ] = \f abc Q_{n+m}^c $.  Note also that when
$\al \neq 0$, $\CY$ and $\CY^*$ are both deformations of the respective
halves of the affine Lie algebra.

The elements $\q 0a$, defined as $t^{(0)} = \sum_a \q 0a s^a$,  generate
an SL(2) subalgebra of $\dy$.  It is possible to extract generators which
are covariant with respect to this SL(2) out of the generators $t^{(n)}$.
For the algebra $\CY$ this is well-known\rdrin.  Here we describe how to
do this for the full double.  Let us first review the result for
$\CY$.  Define
\eqn\IIIBxix{\eqalign{
t^{(0)} &= \sum_{a=1}^3 Q_0^a s^a \cr
t^{(1)} &= \sum_{a=1}^3 Q_1^a s^a + \frac{\al}{4} \q 0a \q 0a . \cr }}
The additional $\al$-dependent term in $t^{(1)}$ is required for the
comultiplication to close.  From \IIIBxviii\ one derives the first equations
in
\IFiv, and from \IIIBxvii\ one obtains \IFi\IFiii.
The deformation of the Serre relations can be proven from
\IIIBxvii\ as follows.  Consider for example $[Q_2^- , Q_1^- ]$.
This can be expressed in terms of the basic generators as
$[Q_2^- , Q_1^- ] = [ Q_1^- , [Q_1^3 , Q_1^- ] ] $.  Expressing this
in terms of $t$'s, one finds
$Q_1^- = 2t_{12}^{(1)} , \q 13 = \t 111 - \t 122$.  Using the relations
\IIIBxvii\ one can thereby evaluate the above commutator:
\eqn\IIIBxx{
[\q 2- , \q 1- ] = - \frac{\al^2}{2}
\( 2 \q 0- \q 13 \q 0- - \q 0- \q 03 \q 1- - \q 1- \q 03 \q 0- \) . }
The latter is equivalent to \IFvi.  The other relations can be proven
similarly.

For the full double $\dy$, the situation is considerably more complicated,
due to the fact that the $t^{(-n)}$, for $n\geq 1$, are infinite series
in the covariant generators.  To see this, consider starting with the
definition
$t^{(-1)} = \sum_a \q {-1}a s^a$.  One finds that
$\Dep (t^{(-1)} )$  does not close on $\Dep (\q {-1} a )$ to any finite
order in $\al$.  One must successively correct $t^{(-1)}$, order by order
in $\al$, in order for $\Dep (t^{(-1)} )$ to close.  Once this is done,
then one finds that the $t^{(-n)}$, $n\geq 2$ can also be expressed in
terms of $\q {-1}a$.  In other words, the algebra $\CY^*$ is generated
solely by the elements $\q {-1}a$, and is a deformation of the universal
enveloping algebra of the negative half of the affine Lie algebra.  We
develop this to lowest orders in $\al$.  Define
\eqn\IIIBxxi{\eqalign{
t^{(-1)} &= \sum_a \q {-1}a s^a + \frac{\al}{4}  \q {-1}a \q {-1}a
 + \ldots
\cr
t^{(-2)} &= \sum_a \q {-2}a s^a + \frac{\al}{4}
\( \q {-1}a \q {-2}a +  \q {-2}a \q {-1}a \) + \ldots .
\cr }}
Then $\Dep (t^{(-1)} ,t^{(-2)} )$ closes to order $\al$ if
\eqn\IIIBxxii{
\Dep (\q {-1}a ) = \q {-1}a \ot 1 + 1 \ot \q {-1}a
+ \al \f abc \q {-1}b \ot \q {-1}c + \ldots , }
where $\q {-2}a \equiv - \f abc \q {-1}b \q {-1}c$.\foot{
The quantum determinant condition equal to one should
also constrain the expansion \IIIBxxi, however we did not
pursue this.}

The relations \IIIBxvii\ imply that
\eqn\IIIBxxiii{
[\q 0a , \q {-1}b ] = \f abc \q {-1}c . }
Other relations obtained from \IIIBxvii\ are deformations of the
affine Lie algebra ones.  In particular, the Serre relations
are deformed\foot{This corrects an error made in  \rls\rsm.}.
One finds
\eqn\IIIBxxiii{
[\q {-2}a , \q {-1}b ] + [\q {-2}b , \q {-1}a ]
= \al^2
\( \f acd \{ \q {-2}b , \q {-2}c , \q {-1}d \}
+ \f bcd \{ \q {-2}a , \q {-2}c , \q {-1}d \} \) +\ldots }
The relation \IIIBxxiii\ is proved as follows.  From \IIIBxxii\ one
can compute the comultiplication of the LHS of \IIIBxxiii\ to
lowest order.  This comultiplication is not consistent with the
RHS equal to zero, but is consistent with the comultiplication
of the RHS of \IIIBxxiii\ to lowest order.

Finally from the relations \IIIBxvii\ one may derive
\eqn\IIIBxxiv{
[\q 1a , \q {-1}b ] = \f abc \q 0c
- \frac{\al^2}{4} \f abc
\( \q {-1}d \q {-1}d \q 0c + \q 0c \q {-1}d \q {-1}d \) + \ldots }
In the formulas \IIIBxxi-\IIIBxxiv\ all of the corrections
are infinite series in $\al$.

\subsec{Remarks on Form Factors}

Form factors are defined as matrix elements of fields on asymptotic
states.  For the field multiplets $\Phi_\Lambda (x)$, they are defined
as
\eqn\IIICi{
f_\La (\th_1 , .. , \th_n ) =
\lvac \Phi_\La (0) \> Z(\th_1 ) \cdots Z(\th_n ) \vac . }
The form factors may thus be understood as Clebsch-Gordon
projectors for the above representations of the affine quantum group.

Smirnov explained how the fundamental axioms of the form factor
bootstrap program may be understood as a consequence of the affine
quantum group symmetry\rsm.  In this section we give a different
derivation of the periodicity axiom (`deformed Knizhnik-Zamolodchikov
equation'), in order to help clarify its meaning.  The equations we
derive improve slightly upon those obtained in \rsm\ by making the
data of the field module $\La$ manifest.  For completeness we
consider the other axioms as well.

It is evident from the exchange relation \IIIBiii\ that $f_\La$ has the
following property upon exchange of two of its vector spaces:
\eqn\IIICii{
f_\La (\th_1 ,..,\th_i , \th_{i+1}, ..,\th_n )
= S_{i,i+1} (\th_{i,i+1} )
f_\La (\th_1 ,..,\th_{i+1} , \th_i, ..,\th_n ) . }

The $2\pi i$-periodicity axiom for the form factors may be derived by
comparing form factors of the fields $\Phi_\La$ and $\Phi'_\La$.
  From general QFT principles (mainly crossing symmetry) one finds that more
general matrix elements can be obtained from the basic ones \IIICi\
as follows \rform:
\eqn\IIICiii{
\langle \th_1 , .., \th_m |
\Phi_\La (0) |\th_{m+1} ,.., \th_n \rangle
= f_\La (\th_1 -i\pi , .., \th_m -i\pi , \th_{m+1},..,\th_n ) .  }
(To simplify the presentation we are dropping transpositions of vector
spaces and charge conjugation matrices in the above notation.)
The field $\Phi'_\La (0)$ is related to $\Phi_\La (0)$ by a rotation
$\Phi'_\La (0) = R_{-\pi} \( \Phi_\La (0) \)$.  This rotation may be
implemented by an operator $U_{-\pi}$:
$\Phi'_\La (0) = U_{-\pi}^\dagger \Phi_\La (0) U_{-\pi} $.  The operator
$U_{-\pi}$ is antiunitarity due to the fact that it involves a time
reversal.  The form factors of $\Phi'_\La (0)$ are thus
\eqn\IIICiv{\eqalign{
\langle \th_1 , .., \th_m |
\Phi_\La' (0) |\th_{m+1} ,.., \th_n \rangle
&=
\langle \th_1 , .., \th_m |
U_{-\pi}^\dagger \Phi_\La (0) U_{-\pi}
|\th_{m+1} ,.., \th_n \rangle \cr
&=
\langle \th_{m+1} + i\pi , .., \th_n + i\pi |
\Phi_\La (0)
| \th_1 + i\pi , .., \th_m + i\pi \rangle \cr
&=
f_\La ( \th_{m+1},..,\th_n , \th_1 + i\pi,..,\th_m + i\pi ) , \cr}}
where we have used the antiunitarity of $U_{-\pi}$ and its  action
on the states.  On the other hand, using the operator formulation
of rotation \IDviii\ one finds
\eqn\IIICv{\eqalign{
\langle \th_1 , .., \th_m |
\Phi_\La' (0) |\th_{m+1} ,.., \th_n \rangle
&=
\langle \th_1 ,.., \th_m | e^a \( \Phi_\La (0) \)
s' (e_a )
 |\th_{m+1} ,.., \th_n \rangle \cr
&=\CR^{-1}_{\rho_{V_{m+1}\cdot\cdot V_n} , \rho_\La}
f_\La (\th_1 -i\pi, .., \th_m -i\pi, \th_{m+1} ,.., \th_n ) . \cr}}
The matrix $\CR_{\rho_{V_{m+1}\cdot\cdot V_n} , \rho_\La }$ is
the universal $\CR$-matrix evaluated in the indicated
representations:
\eqn\rexp{
\CR_{\rho_{V_{m+1} \cdot\cdot V_n }, \rho_\La}
= \CR_{\rho_{V_n} , \rho_\La} \>
\CR_{\rho_{V_{n-1}} ,\rho_\La }\>
\cdots \CR_{\rho_{V_{m+1}} ,\rho_{\La} } . }
Thus, comparing \IIICiv\ and \IIICv, one obtains
\eqn\IIICvi{
f_\La (\th_1 - 2\pi i,..,\th_m -2\pi i , \th_{m+1} ,.., \th_n )
= \CR_{\rho_{V_{m+1} \cdot\cdot V_n} , \rho_\La }
f_\La (\th_{m+1} , .., \th_n ,\th_1 , .., \th_m ) . }
To obtain a periodicity property, one can use \IIICii\ to exchange
the vector spaces.  For example for $m=1$:
\eqn\IIICvi{
f_\La (\th_1 -2\pi i ,\th_2 ,.., \th_n )
= \CR_{\rho_{V_2 \cdot\cdot V_n} , \rho_\La } ~
S_{n1} S_{n-1,1} \cdot\cdot S_{21} ~
f_\La (\th_1 , \th_2 , .., \th_n ) . }

The final form factor axiom may be derived from \IIIBvii,
by simply inserting this expression into \IIICi\ and using
\IIIBxvb \rsm.
As before, we give a version of this formula which displays
the data of the module $\La$:
\eqn\IIICvii{\eqalign{
2\pi i ~
{ \res_{\th_{i+1} = \th_i -i\pi} } ~
& f_\La (\th_1 ,..,\th_i , \th_{i+1} ,..,\th_n ) \cr
&=
\bigl[
\( S_{i,i-1} S_{i,i-2} \cdot\cdot S_{i1} \CR_{V_i ,\rho_\La}
 S_{i,i+2} S_{i,i+3} \cdot\cdot S_{in} \)  -1 \bigr]
\cr
& ~~~\cdot C^{ab} \ket{a}_i \ot \ket{b}_{i+1} ~
f_\La (\th_1 ,\th_2 ,.., \th_{i-1} , \th_{i+2} , .., \th_n ) .\cr}}

When there is no bound state, the relations \IIICii ,
\IIICvi~ and \IIICvii~ are all the form factor axioms. If there is
bound states, an extra residue axiom should be added.

\vfill\eject

\newsec{Discussion}

We conclude this paper with a discussion of some open problems.
The relation \IIICvi\ has been interpreted as a deformed
Knizhnik-Zamolodchikov
 equation\rfr\rsm.  A few remarks are in order.  Consider
the situation where the field is local, so that the factor
$\CR_{\rho_{V_1 \cdot\cdot V_n} , \rho_\La }$ is absent.  Then our
alternative derivation reveals that the relation \IIICvi\ is actually
only a consequence of elementary quantum field theory principles such
as crossing symmetry.  This is really no surprise given the fact that
this property of form factors was proposed some time ago using
bootstrap ideas.  This suggests that this relation is a rather weak
consequence of the affine quantum group symmetry, and that stronger
constraints should exist.

The vertex operators $Z(\th )$ were formally defined by their
intertwining properties.  It should be possible to construct an
explicit representation of these operators in some auxiliary Hilbert
space; e.g. a representation using auxiliary free fields similar
to those used in the vertex operator representations of the quantum
affine algebras\ref\jing{I. B. Frenkel and N. H. Jing,
Proc. Nat'l. Acad. Sci. (USA) 85 (1988) 9373.}.
Such a construction could lead to an efficient means of computing
form factors.

The main difficulty in completely characterizing form factors from the
affine quantum group symmetry is due to the fact that two very
different kinds of representations are needed, the Verma module
field representations $\rho_\La$, and the finite dimensional representations
$\rho_{V_1 \cdot \cdot V_n}$.  The form factors provide an inner product
for these two types of representations.  One would like to establish
a duality between the particle representations and the complete list
of field representations, but it remains unclear whether this is possible.
Such a result would imply that the full Hilbert space of the theory
could be reconstructed from the action of the complete set of fields on the
vacuum.  This is actually the situation encountered in
conformal field theory \ref\bpz{\BPZ}.
Certainly to obtain the complete set of fields will require descending
with local integrals of motion in addition to the non-local ones.
For massive theories such properties are generally very difficult to prove,
since they rely on detailed knowledge of the operator product
expansion, having a deep significance in the realm of
axiomatic quantum field theory.

Finally, let us make an intriguing remark.
It is well known that the classical integrable models are linearized
once they are lifted in the classical double. In view of the fact that
the $S$-matrix is given by the universal $\CR$-matrix and that this
$\CR$-matrix
is the canonical element, i.e. the identity, in $\CA\otimes \CA^*$
embedded in the double $\CD(\CA)\otimes \CD(\CA)$, it is tempting to
conjecture that integrable models admit free realizations once they are
lifted into the quantum double.  This idea may provide a starting
point for constructing vertex operators.

\bigskip\bigskip

\centerline{Acknowledgements}

It is a pleasure to thank O. Babelon, G. Felder and F. Smirnov for discussions.
This work is supported in part by  NATO.
A.L. is supported by the A. P. Sloan Foundation and the National Science
Foundation.

\vfill\eject

\appendix{A}{Quantum Groups}

\def\am{{ $\CA$ }}
\def\a{\CA}
\def\ot{\otimes}
\def\vep{\epsilon}
\def\De{\Delta}

Let \am denote a Hopf algebra.
It is equipped with a multiplication map
$m: \CA \otimes \CA \to \CA$, a comultiplication
$\Delta : \a \to  \a \ot \a$, antipode $s: \a \to \a$, and
counit $\varepsilon : \a \to \Cmath$.
Let $1$ denote the unit element,
 with
$\Delta(1) = 1\ot 1, s(1) = \vep (1) = 1$.  These operations
have the following properties:
\eqna\ai
$$\eqalignno{
m(a\ot 1) &= m(1\ot a) = a &\ai {a} \cr
m(m\ot id ) &= m (id \ot m ) &\ai {b} \cr
(\Delta \ot id ) \Delta &= (id \ot \Delta )\Delta &\ai {c} \cr
\De (a) \De (b) &= \De (ab)  &\ai {d} \cr
m (s\ot id ) \De (a) &= m (id \ot s ) \De (a) = \vep (a) \cdot 1
&\ai {e} \cr
s(ab) &= s(b) s(a) &\ai {f} \cr
\De s &= (s\ot s )   \De^\prime  &\ai {g} \cr
(\vep \ot id ) \De &= (id \ot \vep ) \De = id  &\ai {h} \cr
\vep (ab) &= \vep (a) \vep (b) , &\ai {i} \cr }$$
for  $a,b \in \a$, and $\De^\prime$ is the skew comultiplication
$\De^\prime = P\De$, where $P$ is the permutation operator
$P(a\ot b) = b\ot a$.
Eq. \ai{a} ~ is the definition of the unit element, \ai{b,c}\ are
the associativity and coassociativity of \am , \ai{d} defines
$\De$ to be a homomorphism of \am to $\a \ot \a$, and \ai{e-i}\
are the defining properties of the counit and antipode.
The skew antipode $s'$ is defined by the formula
\eqn\skew{
m(s' \ot id ) \Dep (a) = m (id\ot s' ) \Dep (a) = \vep (a). }

Let $\{ e_a \}$ denote a linear basis for \am .  The above operations
can be formulated by means of structure constants:
\eqna\aii
$$\eqalignno{
e_a \>  e_b &= \m abc \; e_c &\aii {a} \cr
\De (e_a ) &= \u bca \; e_b \ot e_c, ~~~~~
\Dep (e_a ) = \u bca \; e_c \ot e_b   &\aii {b} \cr
s(e_a ) &= s_a^b \; e_b  &\aii {c} \cr
\vep (e_a ) &= \vep_a , &\aii {d} \cr }$$
where $\m abc , \u bca , s_a^b $, and $\vep_a $ are constants
in $\CC$.
We also define constants $\vep^a$ such that
\eqn\aiii{ 1= \vep^a e_a .}

The properties \ai{}\ are easily expressed as consistency
conditions on the structure constants:
\eqna\aiv
$$\eqalignno{
\m abc \> \vep^a &= \m bac \> \vep^a = \delta_b^c &\aiv {a} \cr
\m abc \> \m deb &= \m adb \> \m bec &\aiv {b} \cr
\u bca \> \u deb &= \u dba \> \u ecb &\aiv {c} \cr
\u ija \> \u klb \> \m ikc \> \m jld &= \m abi \> \u cdi &\aiv {d} \cr
\m ijb \> s^j_k \> \u ika &= \m jib \> s^j_k \> \u kia = \vep_a \vep^b
&\aiv {e} \cr
\m abi \> s_i^c &= s_b^j \> s_a^k \> \m jkc &\aiv {f} \cr
\u abi \> s^i_c &= s^a_j \> s^b_k \> \u kjc &\aiv {g} \cr
\u bca \> \vep_c &= \u cba \> \vep_c = \delta^b_a &\aiv {h} \cr
\m abc \> \vep_c &= \vep_a \vep_b &\aiv {i} \cr
}$$

The universal $\CR$-matrix is a construction that will play an important
role in the sequel.  It is defined to satisfy the relation
\eqn\av{
\CR ~ \Dep (e_a ) = \De (e_a ) \CR ,}
for all $e_a$.
The universal $\CR$-matrix may be constructed formally by introducing
the concept of the quantum double, due to Drinfeld.  One defines a dual
Hopf algebra $\CA^*$ with linear basis $\{ e^a \}$, and defines
\eqn\avi{
\CR = \sum_a e_a \ot e^a . }
The properties of $\adu$ are fixed as follows.  The relation \avi\
implies the following relations between the elements $\{ e_a \}$
and $\{ e^a \}$:
\eqn\avii{
\u bca \> \m bid \> e_c \, e^i  = \u bca \> \m icd \> e^i \, e_b  . }
The multiplication map in $\adu$ is fixed by requiring $\CR$ to
satisfy the Yang-Baxter equation:
\eqn\aviii{
\CR_{12} \CR_{13} \CR_{23} = \CR_{23} \CR_{13} \CR_{12} . }
Inserting \avi\ into \aviii, one finds that the Yang-Baxter
equation is satisfied if \avii\ holds along with
\eqn\aix{
e^a \; e^b = \u bac \; e^c . }
The comultiplication in $\adu$ is fixed by imposing the quasi-triangularity
condition
\eqn\ax{
(1\ot \De ) \> \CR_{12} = \CR_{13} \CR_{12} ,
}
which implies
\eqn\axi{
\De (e^a ) = m^a_{bc} \; e^b \ot e^c . }
Using the properties \aiv{}, one may define the antipode and counit
such that \ai{e-i} are satisfied in $\adu$:
\eqn\axii{
s(e^a) = (s')^a_b e^b,  ~~~~~\vep (e^a ) = \vep^a ,~~~~~\vep_a e^a = 1.}
One easily checks that the above definitions indeed endow $\adu$ with
a Hopf algebra structure by using the properties \aiv{} of the
structure constants.  The Hopf algebra $\CA \ot \adu$ is referred to
as the quantum double $\CD (\CA )$.
One also has
\eqn\axiii{
\CR^{-1} = \sum_a s' (e_a ) \ot e^a . }

The relations \avii\ between elements of $\CA$ and $\adu$ can also be
expressed in the form
\eqna\axiv
$$\eqalignno{
e^n \, e_m  &=
\u bca \u afm \m bid \m djn s^j_f ~ e_c \,  e^i
&\axiv {a} \cr
e_m \, e^n &=
\u bca \m icd \u fam \m jdn s^j_f ~ e^i \, e_b .
  &\axiv {b} \cr
}$$
The formula \axiv{a}\ is established by multiplying both sides
of \avii\ by $\u afm \m djn s^j_f$  and using \aiv{b,c,e,a,h} .
Equation \axiv{b} is similarly proven.

\appendix{B}{Verifying the Current Representation of $\CD (\CA)$.}

In this appendix we prove that the current representation
$\rho_{{}_J}$ defined in \IEi\ and \IEv\ satisfy the defining
relations of $\CD (\CA )$.

By construction, namely due to the property \IBxii\ of the
adjoint action, $\rho_{{}_J}$ satisfies the multiplication law in
$\CA$:
\eqn\bi{
\rho_{{}_J} (e_a)_i^d \> \rho_{{}_J} (e_b)_c^i
= m_{ab}^j ~ \rho_{{}_J}  (e_j)_c^d .}
The coassociativity condition \aiv{c}\ ensures that
$\rho_J (e^a)$ satisfies the multiplication law in $\adu$:
\eqn\bii{
\rho_{{}_J}  (e^a)_i^d \> \rho_{{}_J}
(e^b)_c^i = \mu^{ba}_i \> \rho_{{}_J}  (e^i )_c^d . }

That the current representation also satisfies the defining relation
\IDvi\ between elements of $\CA$ and $\adu$ is more non-trivial,
and may be proven as follows.  Inserting the current representation
into \IDvi, one finds these latter relations are satisfied if
\eqn\biii{
\mu^{bc}_a m_{bi}^d \mu^{il}_m \tilde{f}^n_{cl}
= \mu^{bc}_a m_{ic}^d \tilde{f}^l_{bm} \mu^{in}_l . }
An indirect proof of the identity \biii\ can be given as follows.
The $\De$-homomorphism property \ai{d} applied to the relation
\IBxv, namely, requiring $\De (Q_a)$ to also satisfy \IBxv, leads to
the formula
\eqn\biv{
Q_a \Th_b^c + f^c_{ij} \Th^i_a \Th ^j_b = f^d_{ab} \Th^c_d
+ R^{ed}_{ab} \Th^c_d Q_e .}
Evaluating \biv\ in the adjoint representation\IBxiii\  by using
\IBxo\ and  \IBxvii\ yields directly the identity \biii.

\vfill\eject

\figures
\fig{1}{Graphical representation of non-local currents.}
\fig{2}{Action of $\Th_a^b$ on a field.}
\fig{3}{Contour decomposition of the adjoint action.}
\fig{4}{Contour decomposition leading to the comultiplication.}
\fig{5}{$R_\pi^2$ on a non-local current.}
\fig{6}{Euclidean rotation used to compute the antipodes.}
\fig{7}{$ad'$ action.}
\fig{8}{Graphical representation of eq. \IEix.}

\listrefs
\bye